\documentclass[twocolumn,preprintnumbers,prd,superscriptaddress]{revtex4-1}

\usepackage{graphicx}
\usepackage{dcolumn}
\usepackage{bm}

\usepackage[colorlinks=true,linkcolor=black, citecolor=black,
urlcolor=black]{hyperref}

\usepackage{relsize}
 \usepackage{multirow,graphics}
 \usepackage{amstext}
 \usepackage{amssymb}
 \usepackage{amsmath}
 \usepackage{mathfixs}
 \usepackage{graphicx}
 \usepackage{color}
 \usepackage{dsfont} 
 \usepackage{delimset} 

 \usepackage[caption=false]{subfig}

\usepackage{xcolor}

\newcommand{\sfrac}[2]{{\textstyle\frac{#1}{#2}}}
\newcommand{\half}{\sfrac{1}{2}}
\newcommand{\ihalf}{\sfrac{i}{2}}

\newcommand{\gen}[1]{\mathrm{#1}}

\newcommand{\levz}{\mathrm{J}}
\newcommand{\dd}{\mathrm{d}}

\newcommand{\alg}[1]{\mathfrak{#1}}

\newcommand{\STconst}{X}
\newcommand{\PDE}{\mathrm{PDE}}
\newcommand*\FF[4]{F_{#1}\big[\genfrac{}{}{0pt}{1}{#2}{#3};#4\big]}
\newcommand{\limitstack}[3]{{\substack{{\scriptscriptstyle #1}\\ {\scriptscriptstyle #2}\\{\scriptscriptstyle #3}}}}

\makeatletter
\newlength{\apb@width}
\newcommand{\autoparbox}[2][c]{\settowidth{\apb@width}{#2}\parbox[#1]{\apb@width}{#2}}
\newcommand{\includegraphicsbox}[2][]{\autoparbox{\includegraphics[#1]{#2}}}
\makeatother

\makeatletter
\def\mr@ignsp#1 {\ifx\:#1\@empty\else #1\expandafter\mr@ignsp\fi}%
\newcommand{\multiref}[1]{\begingroup
\xdef\mr@no@sparg{\expandafter\mr@ignsp#1 \: }%
\def\mr@comma{}%
\@for\mr@refs:=\mr@no@sparg\do{\mr@comma\def\mr@comma{,}\ref{\mr@refs}}%
\endgroup}
\makeatother

\newcommand{\hypref}[2]{\ifx\href\asklfhas #2\else\href{#1}{#2}\fi}
\newcommand{\Secref}[1]{section~\multiref{#1}}
\newcommand{\Appref}[1]{appendix~\multiref{#1}}

\newcommand{\Tabref}[1]{table~\multiref{#1}}

\newcommand{\Figref}[1]{figure~\multiref{#1}}

\renewcommand{\eqref}[1]{(\multiref{#1})}

\begin{document}

\preprint{HU-EP-19/39}

\title{Yangian Bootstrap for Conformal Feynman Integrals}

\author{Florian Loebbert}
\email{loebbert@physik.hu-berlin.de} 
\affiliation{%
Institut f\"ur Physik, Humboldt-Universi\"at zu Berlin,
Zum Gro{\ss}en Windkanal 6, 12489 Berlin, Germany
}%
\author{Dennis M\"uller} 
\email{dennis.mueller@nbi.ku.dk}
\affiliation{%
Niels Bohr Institute, Copenhagen University, Blegdamsvej 17, 
2100 Copenhagen, Denmark}%
\author{Hagen M\"unkler} 
\email{muenkler@itp.phys.ethz.ch}
\affiliation{%
Institut f\"ur Theoretische Physik,
Eidgen\"ossische Technische Hochschule Z\"urich,
Wolfgang-Pauli-Strasse 27, 8093 Z\"urich, Switzerland}%



\date{\today}

\begin{abstract}
We explore the idea to bootstrap Feynman integrals using integrability. In particular, we put the recently discovered Yangian symmetry of conformal Feynman integrals to work.
As a prototypical example we demonstrate that the $D$-dimensional box integral with generic propagator powers is completely fixed by its symmetries to be a particular linear combination of Appell hypergeometric functions. In this context the Bloch--Wigner function arises as a special Yangian invariant in 4D. The bootstrap procedure for the box integral is naturally structured in algorithmic form. We then discuss the Yangian constraints for the six-point double box integral as well as for the related hexagon.  For the latter we argue that the constraints are solved by a set of generalized Lauricella functions and we comment on complications in identifying the integral as a certain linear combination of these. Finally, we elaborate on the close relation to the Mellin--Barnes technique and argue that it generates Yangian invariants as sums of residues. 
\end{abstract}

                              
\maketitle

%
\section{Introduction}\label{sec:intro}

Theoretical predictions for particle phenomenology strongly depend on our understanding of Feynman integrals. When the number of loops and legs increases, computations quickly become intractable. 
Facing these problems, theorists are challenged to identify new methods to evaluate these integrals and to unveil their deeper mathematical structure.
Recently a new infinite dimensional Yangian symmetry was identified for a large class of so-called fishnet Feynman graphs \cite{Chicherin:2017cns,Chicherin:2017frs}. In the present paper we explore this connection between Feynman integrals and the theory of integrable models, which play a crucial role for developing analytical methods in all areas of physics. Notably, these scalar fishnet integrals furnish some of the most important building blocks of quantum field theory at any loop order. Their integrability properties can be understood through their interpretation as correlation functions of an integrable bi-scalar fishnet model, which represents an elegant reduction of deformed $\mathcal{N}=4$ super Yang--Mills theory  \cite{Gurdogan:2015csr}. Via this relation the integrability features of the AdS/CFT correspondence find their way to phenomenologically relevant building blocks of generic quantum field theories. Moreover, this makes connection to an alternative interpretation of fishnet integrals in terms of integrable vertex models, which was discovered by Zamolodchikov almost fourty years ago~\cite{Zamolodchikov:1980mb}.

In the present paper we investigate the constraining power of the Yangian for conformal Feynman integrals. In particular, we discuss the respective constraints for the first two non-trivial cases of fishnet graphs in four dimensions. These are the completely off-shell one-loop box and the two-loop double box integral
\footnote{Though we will work in the dual momentum space throughout this paper, we refer to the respective Feynman diagrams in the original momentum space, i.e.\ we speak of the box instead of the cross integral. The hat over the integral symbol $\hat I$ denotes the integral with undeformed propagator powers as opposed to the symbol $I$ used later for generic propagator powers.
}:
\begin{equation}
\hat I_4=
\includegraphicsbox[scale=.7]{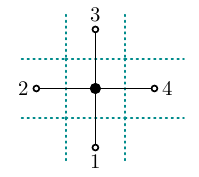}
\quad
\hat I_{3,3}=
\includegraphicsbox[scale=.7]{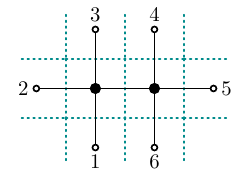}
\end{equation}
Conformal symmetry allows to write these integrals as
functions of 2 and 9 cross ratios, respectively. The case of the double box integral is particularly interesting since it has not been solved so far. Recently, a lot of progress was made on understanding the 7-cross-ratio limit of this integral, for which the two middle legs in momentum space (dotted lines) are put on shell 
\cite{Bourjaily:2017bsb,Adams:2018bsn}. In this limit the integral is known to be described by elliptic functions \cite{CaronHuot:2012ab}. Due to its interesting relation to the double box \cite{Paulos:2012nu} as outlined below, we also discuss the 9-variable hexagon integral:
\begin{equation}
\hat I_6=\includegraphicsbox[scale=.7]{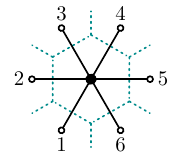}
\end{equation}
Also for this integral results are only known in a three-particle on-shell case resulting in a function of 6 variables \cite{Dixon:2011ng,DelDuca:2011wh}.

The Yangian symmetry employed in this paper provides the algebraic foundation of rational integrable models.
Traditionally it appears as a symmetry of integrable S-matrices in two dimensions, where it typically fixes the scattering matrix completely, cf.\ \cite{Loebbert:2018lxk}. One may thus expect similarly strong constraints for the above box, double box and hexagon integrals.

In the following we show that indeed the Yangian can be used to fix the $D$-dimensional box integral with generic propagator powers. 
We then discuss the analogous constraints for the double box and the related hexagon integral. These constraints are formulated as systems of differential equations in the conformal cross ratios for the respective Feynman integrals. For the hexagon we argue that the Yangian constraints are solved by a large set of generalized (Srivastava--Daoust) Lauricella series in 9 variables, whose exact domains of convergence remain unclear. We discuss a recursive strategy to fix overall constants of the considered integrals by relating them to the star-triangle equation in coincidence limits of external points. Finally, we comment on the close relation of this bootstrap approach to the Mellin--Barnes technique and the common convergence issues faced for the considered six-point integrals. We close with an extended outlook pointing at various promising future directions.

\section{Conformal Yangian}

Conformal Feynman integrals in $D$ dimensions are built from $n$-point vertices such that the powers $a_j$ of the $n$ connected propagators obey $\sum_{j=1}^n a_j =D$, e.g.\ at one loop we have
\begin{equation}
\int  \frac{\dd^D x_0}{x_{10}^{2a_1}\dots x_{n0}^{2a_n}} =
\includegraphicsbox[scale=.8]{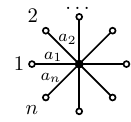}
\label{eq:OneLoopStar}
\end{equation}
Integrals built from such vertices are conformal, i.e.\ they transform covariantly
under the differential generators $\levz^A$ of the conformal Lie algreba $\alg{so}(1,D+1)$, whose densities read
\footnote{Note that by inclusion of a non-trivial numerator factor we could make the integrals conformally invariant.}
\begin{align}\label{eq:lev0gens}
 \gen{P}_\mu& =-i\partial_\mu,
&
\gen{ L}_{\mu\nu}&=ix_\mu \partial_\nu-ix_\nu \partial_\mu , 
\\
 \gen{D}&=-i x_\mu \partial^\mu-i\Delta,
&
\gen{K}_\mu&= 2  x^{\nu}  L_{\nu \mu} - i x^2  \partial_{\mu} - 2 i \Delta x_{\mu}.\nonumber
 \end{align}
Here the conformal dimension $\Delta$ has to reflect the weight of the respective integral, e.g.\ for the above one-loop integral \eqref{eq:OneLoopStar} one sets $\Delta_j=a_j$ for $j=1,\dots,n$.
Due to their conformal symmetry, these integrals can be written in the form
 \begin{equation}
 \label{eq:ConfInvariant}
 I_n = V_n \,\phi(u_1,\dots, u_N).
 \end{equation}
 Here the prefactor $V_n$ carries the conformal weight of the integral while the variables $u_j$ denote the conformal cross ratios whose number $N$ depends on the number $n$ of external points.

For $n=3$ it is not possible to construct conformal cross ratios and hence, the above function $\phi$ is constant. This is reflected in the well known \emph{star-triangle} or \emph{uniqueness relation}, which holds for the conformality condition $a+b+c=D$:
\begin{equation}
\label{eq:StarTriangle}
\includegraphicsbox[scale=.8]{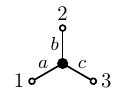}
\hspace{-3mm}=
\int  \frac{\dd^D x_0}{x_{10}^{2a}x_{20}^{2b}x_{30}^{2c}}
=
 \frac{\STconst_{abc}}{x_{12}^{2c'}x_{23}^{2a'}x_{31}^{2b'}}
=\STconst_{abc}
\hspace{-3mm}\includegraphicsbox[scale=.8]{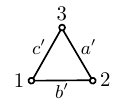}  
\end{equation}
Here we have defined $a'=D/2-a$ as well as 
\begin{equation}
\STconst_{abc}
:=\pi^{D/2}\frac{\Gamma_{a'}\Gamma_{b'}\Gamma_{c'}}{\Gamma_{a}\Gamma_{b}\Gamma_{c}},
\end{equation}
see e.g.\ \cite{Chicherin:2012yn}, with $\Gamma_x=\Gamma(x)$ denoting the Gamma function.

At four points, the function $\phi$ in \eqref{eq:ConfInvariant} becomes non-trivial due to the presence of two non-trivial conformal invariants $z$ and $\bar z$ defined by
\begin{equation}
z\bar z=\frac{x_{12}^2 x_{34}^2}{x_{13}^2 x_{24}^2}, 
\qquad
(1-z)(1-\bar z)=\frac{x_{14}^2 x_{23}^2}{x_{13}^2 x_{24}^2}. 
\label{Def:zzbar}
\end{equation}
Hence, conformal symmetry is no longer sufficient to completely fix the four-point function and it is natural to ask how one can further constrain the conformal four-point integral
\begin{equation}
\hat I_4=\frac{1}{{x_{13}^{2}x_{24}^{2}}}{\hat\phi(z,\bar z)}.
\end{equation} 
As noted in the introduction, the class of fishnet graphs was recently shown to feature an infinite dimensional extension of the conformal Lie algebra \cite{Chicherin:2017cns,Chicherin:2017frs}. The so-called Yangian algebra is generated by the level-zero generators given in \eqref{eq:lev0gens} and the level-one generators taking the form
 \begin{equation}\label{eq:lev1}
 \gen{ \widehat J}^A = \half f^{A}{}_{BC} \sum_{k=1}^n\sum_{j=1}^{k-1}\gen{J}_j^C\gen{J}_k^B +\sum_{j=1}^n s_j\gen{J}_j^A.
 \end{equation}
Here $f^{A}{}_{BC}$ denote the dual structure constants of the conformal algebra $\alg{so}(1,D+1)$ and the so-called evaluation parameters $s_j$ are numbers associated to each Feynman graph, cf.\ \cite{Chicherin:2017cns} and \Secref{sec:EvalParam}.
For instance, the level-one momentum generator reads
\begin{equation}\label{eq:Phat}
\gen{\widehat P}^{\mu}=-\ihalf\sum_{j<k=1}^n\big[(\gen{L}_j^{\mu\nu}+\eta^{\mu\nu}\gen{D}_j)\gen{P}_{k,\nu}-(j\leftrightarrow k)\big]
+\sum_{j=1}^n s_j \gen{P}_j^\mu.
\end{equation}
Notably, these level-one generators act non-locally on the external legs of the Feynman graphs, i.e.\ they have a non-trivial coproduct
 \footnote{In addition the definition requires the so-called Serre relations, see e.g.\ \cite{Loebbert:2016cdm} for more details. Invariance under the full level-zero (conformal) algebra and a single level-one generator guarantees invariance under the full Yangian algebra.}.
The resulting invariance equations can be translated into partial differential equations (PDEs) in the conformal cross ratios. 
More explicitly, the application of the level-one generator to the integral $I_n$ yields
\begin{equation}
0 = \gen{\widehat P}^{\mu}  I_n = V_n \sum_{j<k=1}^n \frac{x_{jk}^\mu}{x_{jk}^2} \, \mathrm{PDE}_{jk} \phi .  
\label{eq:PhatonIn}
\end{equation}
Here, the coefficients $\mathrm{PDE}_{jk}$ denote differential operators depending only on the cross ratios
and we can employ conformal transformations 
in order to vary their prefactors independently. As shown in \Appref{sec:CrossRatios} the above Yangian invariance condition 
requires that 
\footnote{Here one may wonder how the 15 degrees of freedom of the conformal group can lead to 15 invariance conditions, 
when it is clear that translations and dilatations do not contribute. The point is that the conformal group 
acts non-linearly on the $x_{ij} ^\mu / x_{ij}^2$ such that the linear counting of degrees of freedom does not work out.}
\begin{align}
\mathrm{PDE}_{jk} \phi = 0,
 \qquad 1\leq j<k\leq n , 
 \label{eq:PDEeq0}
\end{align}
at least as long as we have no more than six external points. 
Notably, this makes connection between the Yangian symmetry of fishnet Feynman graphs
and systems of differential equations for Feynman integrals which have been studied in various contexts. 
We will exploit these Yangian differential equations in the following.

\section{Bloch--Wigner Function from Yangian Symmetry}
\label{sec:BWfromY}

In order to illustrate the constraining power of the Yangian algebra we start by considering the one-loop box integral for the special case of propagator weights $a_j=1$ in $D=4$ dimensions:
 \begin{equation}
 \label{eq:4ptUndeformed}
 \hat I_4=\int  \frac{\dd^4 x_0}{x_{10}^{2}x_{20}^{2}x_{30}^{2} x_{40}^{2}}
=
\includegraphicsbox[scale=.8]{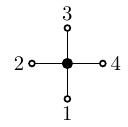}
 \end{equation}
Writing $\hat I_4={\hat\phi(z,\bar z)}/{x_{13}^{2}x_{24}^{2}}$,
the Yangian invariance of the box integral translates into a system of two partial differential equations
\begin{align}\label{eq:BoxPDEsUnsymz}
0=[D_j(z)-D_j(\bar z)]\hat \phi(z,\bar z),
\quad
j=1,2,
\end{align}
where the differential operators $D_j$ are given by
\begin{align}
D_1(z)&=
z (z-1)^2 \partial_z^2 +(3 z-1) (z-1)
  \partial_z+z,
\\
D_2(z)&=
z^2(z-1)  \partial_z^2 +(3 z-2) z
   \partial_z+z.
\end{align}
Clearly, a solution to the first differential equation will not automatically solve the second equation and vice versa. 
It is thus natural to ask which boundary conditions lead to simultaneous solutions. We consider boundary conditions 
on the line $z=\bar z$, which is the natural boundary of the kinematic space described by the $x_i$, cf.\ 
\Appref{sec:CrossRatios} for a more detailed discussion. 
Expanding the equations around this boundary, we find that the combination of both differential equations 
constrains the boundary conditions to four possible functions. 

In order to find the complete solution of the above system, we introduce the coordinates $z_1$ and $z_2$ as the real and imaginary part of $z= z_1 + i z_2 $, and
we expand the equations around
a generic point $a=(z-\bar z)/2 i$. 
Moreover, we introduce the function $\psi(z_1,z_2)=z_2\,\hat \phi (z_1,z_2)$ and expand around the line $z_2=a$:
\begin{align}
\psi (z_1,z_2) = \sum \limits _{n=0} ^\infty \frac{(z_2-a)^n}{n!} f_{a,n}(z_1) . 
\end{align}
Note that this form is completely general (for generic $a$), since we do not need to consider the 
possibility that $\hat\phi$ diverges for generic $a$.
The above differential equations \eqref{eq:BoxPDEsUnsymz} now translate into differential equations for the coefficient functions $f_{a,n}$. In particular, the full solution for the above box integral can be obtained from $ f_{a,0}$ via the relation $\psi (z_1,a) = f_{a,0}(z_1)$. Here $f_{a,0}$ is essentially found by solving an ordinary third-order differential equation, which can be done straightforwardly in Mathematica. 
The integration constants appearing in the solution of these ordinary differential equations can be fixed, e.g.\ 
by requiring that
\begin{align}
\partial_{a} f_{a,0} (z_1) = f_{a,1} (z_1) . 
\end{align}
In agreement with the boundary conditions, the full solution to the Yangian constraints obtained in this way has four free parameters~$c_j$:
\begin{align}
\hat \phi(z,\bar z)= \sum_{j=1}^4 c_j \frac{f_j(z,\bar z)}{z-\bar z}.
\end{align}
Here we have defined
\begin{align}
f_1&=1 ,
\\
f_2&=\log(\bar z)-\log (z ), 
\\
f_3&=\log(1-\bar z) -\log (1- z),
\\
f_4&=2 \text{Li}_2(z) - 2 \text{Li}_2(\bar z) 
		+ \log \sfrac{1-z}{1-\bar z} \log (\bar z z) .
		\label{eq:BW}
\end{align}
Obviously, the box integral is invariant under permutations of any of its external legs. This results in functional relations for $\hat\phi(z,\bar z)$:
\begin{align}
\hat \phi(z,\bar z)&=\hat\phi\brk!{1-z,1-\bar z},
\label{eq:FuncRelBox1}
\\
z \bar z \hat \phi(z,\bar z)&=\hat\phi\brk!{z^{-1},\bar z^{-1}}.
\label{eq:FuncRelBox2}
\end{align}
Imposing these permutation symmetries on $\hat\phi$ uniquely fixes the solution to be given by the well known Bloch--Wigner function $f_4$ given in \eqref{eq:BW} divided by an overall factor $z-\bar z$:
\begin{equation}
\label{eq:BWsolution}
\hat \phi(z,\bar z)=c_4 \frac{f_4(z,\bar z)}{z-\bar z}.
\end{equation}
Below we will also demonstrate that the overall constant is fixed by the star-triangle integral \protect\eqref{eq:StarTriangle}
and takes the value $c_4=\pi^2$.
This is in agreement with the results in the literature~\cite{Usyukina:1992jd}.

In conclusion, the four-point box integral is completely fixed by its symmetries. 
Note that we did not \emph{assume} any boundary conditions, nor did we use an ansatz to obtain the solution. The situation resembles the star-triangle relation at three points, which is fixed by the level zero of the Yangian, i.e.\ by the conformal Lie algebra symmetry, see \Tabref{tab:STvs4pt}.
\begin{table}[t]
\begin{center}
 \renewcommand{\arraystretch}{1.2}
\begin{tabular}{|l|c|c|}\hline
Yangian &3 points & 4 points\\\hline
level zero& fixed &$\hat \phi(z,\bar z)$\\[-1pt]
& Star-Triangle&\\\hline
level one $+$ perm.\  &---& fixed\\[-1pt]
& & Bloch--Wigner\\\hline
\end{tabular}
\end{center}
\caption{The star-triangle integral \protect\eqref{eq:StarTriangle} is fixed by the level-zero Yangian symmetry, i.e.\ by invariance under the conformal Lie algebra generators \protect\eqref{eq:lev0gens}. Similarly, the four-point integral \protect\eqref{eq:4ptUndeformed} is fixed by invariance under the level-one Yangian generators \protect\eqref{eq:lev1} supplemented by permutation symmetries. }
\label{tab:STvs4pt}
\end{table}

\section{Parametric Box in \texorpdfstring{$D$}{D} Dimensions}
\label{sec:ParamBox}

Next we would like to understand more generic Yangian-invariant four-point functions. A natural extension is to generalize the above four-point box to $D$ spacetime dimensions and to introduce generic propagator powers:
 \begin{equation}
 \label{eq:BoxIntD}
 I_4= \int  \frac{\dd^D x_0}{x_{10}^{2a}x_{20}^{2b}x_{30}^{2c} x_{40}^{2d}}
=
\includegraphicsbox[scale=.8]{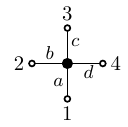}
 \end{equation}
 Conformal symmetry requires that $a+b+c+d =D$
 and that the scaling dimensions entering \eqref{eq:lev0gens} take values
 \begin{equation}
 \Delta_j=(a,b,c,d)_j,
 \qquad
 j=1,\dots,4.
 \end{equation}
Note that using the star-triangle relation \eqref{eq:StarTriangle}, this integral can be mapped (modulo an external propagator) to a two-loop integral 
with two connected three-point stars:
\begin{equation}
\includegraphicsbox[scale=.8]{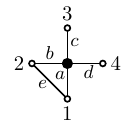}
=
\STconst^{-1}_{a'b'e'}
\includegraphicsbox[scale=.8]{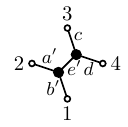}
 \end{equation}
Here the parameter $e$ is fixed through the constraint $a+b+e=D/2$. Note that the propagators on the right
hand side sum up to $D$ at each of the two integration vertices. For $D=6$ this integral is the natural four-point
Yangian invariant composed of three-point vertices and with propagator weights
\begin{equation}
a=1,
\quad
b=1,
\quad
c=2,
\quad
d=2,
\quad
e=1.
\end{equation}
To investigate the Yangian invariance of the above $D$-dimensional integral with generic propagator powers we write $I_4= V_4\,\phi(u,v)$, where
\begin{equation}
V_4=x_{14}^{2b+2c-D}x_{13}^{2d-D}x_{34}^{-2c-2d+D}x_{24}^{-2b}, 
\end{equation}
and note that the evaluation parameters for the Yangian generators are given 
in equation \eqref{eq:BoxEvalParam}. 
For conciseness we introduce the Euler operators $\theta_j= v_j\partial_{v_j}$ with 
\begin{align}
u\equiv v_1&
=z\bar z, 
\\
v\equiv v_2&
=(1-z)(1-\bar z),
\end{align}
and the shorthand $\theta_{jk}=\theta_j+\theta_k$. The Yangian constraints then translate into the following parametric differential equations:
\begin{align}
0&=\brk*{\alpha \beta +(\alpha+\beta)\theta_{12}+\theta_{12}^2-\theta_{1}\partial_{u} -\gamma \partial_{u}}\phi(u,v),
\nonumber\\
0&=\brk*{\alpha \beta +(\alpha+\beta)\theta_{12}+\theta_{12}^2-\theta_{2}\partial_{v} -\gamma' \partial_{v}}\phi(u,v).
\label{eqn:AppellPDEsEuler}
\end{align}
Here greek parameters are given in terms of latin propagator powers and the spacetime dimension $D$:
\begin{align}
\alpha&=b,
&
\gamma&=+\sfrac{D}{2}-c-d+1,
\nonumber\\
\beta&=\sfrac{D}{2}-d,
&
\gamma'&=-\sfrac{D}{2}+b+c+1.
\label{eq:RelsGreekLatin}
\end{align}
Importantly, equations \eqref{eqn:AppellPDEsEuler} can be identified with the system of partial differential equations defining the \emph{Appell hypergeometric function} $F_4$ of two variables $u$ and $v$ 
\footnote{Note that the Appell function $F_4$ is symmetric in $\alpha$ and $\beta$.}:
\begin{equation}
\FF{4}{\alpha,\beta}{\gamma,\gamma'}{u,v} = \sum\limits_{m,n=0}^{\infty} \frac{(\alpha,m+n)(\beta,m+n)}{(\gamma,m)(\gamma',n)(1,m)(1,n)} \, u^m v^n \, ,
\label{eqn:F4Def}
\end{equation}
Here the Pochhammer symbol is given by the ratio of Gamma functions
$
(\lambda,k)= \Gamma_{\lambda+k}/\Gamma_\lambda.
$
In agreement with our findings on the special case in the previous section, it is well known that the space of solutions to the above PDEs is spanned by four functions \cite{appell1926fonctions}:
\begin{align}
g_1&= \FF{4}{\alpha,\beta}{\gamma,\gamma'}{u,v},
\label{eq:g1}
\\
g_2&= u^{1-\gamma}\FF{4}{\alpha+1-\gamma,\beta+1-\gamma}{2-\gamma,\gamma'}{u,v},
\\
g_3&= v^{1-\gamma'}\FF{4}{\alpha+1-\gamma',\beta+1-\gamma'}{\gamma,2-\gamma'}{u,v},
\\
g_4&= u^{1-\gamma}v^{1-\gamma'}\FF{4}{\alpha+2-\gamma-\gamma',\beta+2-\gamma-\gamma'}{2-\gamma,2-\gamma'}{u,v}.
\label{eq:g4}
\end{align}

The final steps for fixing this integral will be outlined in detail in \Secref{sec:recsBox}. For completeness let us already note that we can employ the permutation symmetries of the box integral to completely fix the solution up to an overall constant $N_4$:
\begin{align}
&\phi(u,v)=N_4
\Big[\Gamma_\alpha\Gamma_{\beta} \Gamma_{1-\gamma'} \Gamma_{1-\gamma}\, 
g_1(u,v)
 \label{eqn:YDefCrossF4Final} \\
&+\, \Gamma_{1+\alpha-\gamma} \Gamma_{1+\beta-\gamma}\Gamma_{\gamma-1}\Gamma_{1-\gamma'} \, 
g_2(u,v)
\nonumber\\
&+ \, \Gamma_{1+\alpha-\gamma'} \Gamma_{1+\beta-\gamma'} \Gamma_{1-\gamma}\Gamma_{\gamma'-1} \, 
g_3(u,v)
\nonumber \\
&+ \, \Gamma_{2+\beta-\gamma-\gamma'}  \Gamma_{2+\alpha-\gamma-\gamma'}\Gamma_{\gamma'-1}\Gamma_{\gamma-1}\, 
g_4(u,v)\Big]\, .\nonumber
\end{align}
The overall constant can be fixed by comparison with the star-triangle integral in a coincidence limit of two external points: 
\begin{equation}
N_4= \frac{\pi ^{2+\alpha+\beta-\gamma-\gamma'}}
	{\Gamma_\alpha \Gamma_{1+\beta-\gamma}\Gamma_{1+\beta-\gamma'}\Gamma_{2+\alpha-\gamma-\gamma'}} . 
\label{eqn:BoxNormalizConst}
\end{equation}

If we send one of the external points of the above four-point invariant to infinity via a conformal transformation, this result perfectly agrees with the triangle integral computed by Boos and Davydychev \cite{Boos:1990rg}.

As already pointed out in the classic reference \cite{appell1926fonctions}, the limit $a,b,c,d\to 1$ of unit propagator powers in $D=4$ is subtle, since in this limit the above four solutions \eqref{eq:g1}--\eqref{eq:g4} coincide. Moreover, their coefficients in \eqref{eqn:YDefCrossF4Final} diverge. Careful investigation shows that the solution is given by, cf.\ \cite{Davydychev:1992xr}:
\begin{align}
\phi(u,v)=\sfrac{\pi^4}{3} h_1(u,v)+ \pi^2 h_2(u,v) \, .
\label{eqn:BoxZeroDefF4}
\end{align}
Here $h_1(u,v)=F_4(1,1,1,1,u,v) $ and using the notation
$f_{,\alpha}:= \partial_\alpha f$ and $f_{,\alpha\beta}:= \partial_\alpha \partial_\beta f$ we have defined
\begin{align} 
&h_2=\big[h_1 \log(u) \log(v) +\log(u)\left( F_{4,\alpha}+ F_{4,\beta}+2 F_{4,\gamma'} \right)
\nonumber\\
&+\log(v)\left( F_{4,\beta}+ F_{4,\beta}+2F_{4,\gamma}\right) \nonumber 
+ F_{4,\alpha\alpha}+F_{4,\beta\beta}+2F_{4,\alpha\beta}\\
&+2F_{4,\alpha\gamma}+2F_{4,\alpha\gamma'}+2 F_{4,\beta\gamma} +2F_{4,\beta\gamma'}+4F_{4,\gamma\gamma'} \big]_{
\genfrac{}{}{0pt}{2}{\alpha, \beta}{\gamma,\gamma'} =1 } \, . \nonumber 
\end{align}
This result indeed reproduces the Bloch--Wigner function \eqref{eq:BWsolution} as found above.

\section{Bootstrapping the Box}
\label{sec:recsBox}

In this section we demonstrate explicitly how to bootstrap the box integral with generic propagator powers from scratch.
This is particularly instructive in view of the more involved examples considered in the subsequent sections.
In order to solve the Yangian differential equations, we make a power series ansatz
\begin{equation}
\phi (u, v) = 
\sum \limits _{m,n} 
 g_{mn}^{\alpha  \beta  \gamma  \gamma'}  u^{m} v^{n}, 
\end{equation}
and translate the PDEs into
the following set of recurrence relations for the coefficient functions $g_{m,n}^{\alpha  \beta  \gamma  \gamma'}$:
\begin{align}
g_{m,n+1}^{\alpha  \beta  \gamma  \gamma'}  &= 
	\frac{(m + n + \alpha)(m + n + \beta)}
		{(n + 1)(n + \gamma')} g_{mn} ^{\alpha  \beta  \gamma  \gamma'} , 
\label{eqn:Rec-1} \\
g_{m+1,n}^{\alpha  \beta  \gamma  \gamma'}  &=  
	\frac{(m + n + \alpha)(m + n + \beta)}
		{(m + 1)(m + \gamma)} g_{mn} ^{\alpha  \beta  \gamma  \gamma'} .  
\label{eqn:Rec-2}
\end{align}
These are straightforwardly solved using Mathematica and the solution can be brought to the following form
\footnote{Note that the identity 
\protect\eqref{eq:ReflId2} 
given in the appendix can be useful to
bring the solution into this form.}, which is of course only determined up to an overall constant:
\begin{align}
g_{mn} ^{\alpha  \beta  \gamma  \gamma'}  = 
	\frac{1}{\Gamma_{m+1}\Gamma_{n+1} \Gamma_{m+\gamma}
	\Gamma_{n+\gamma'} \Gamma_{1-m-n-\alpha}\Gamma_{1-m-n-\beta}} .
\label{eqn:Rec-sol-1}
\end{align}
We will refer to this expression as the \emph{fundamental solution}.
 Note that in order to show that $g_{mn}$ solves the 
above recurrence equations, it is not necessary to assume that $m,n$ are integers.
We have hence found a formal solution of the PDEs for every $x,y \in [0,1)$: 
\begin{align}
\label{eq:SolutionBasisBox}
G_{x,y} ^{\alpha  \beta  \gamma  \gamma'} (u, v) = 
	\sum \limits _{\genfrac{}{}{0pt}{1}{m \in x + \mathbb{Z}}{n \in y + \mathbb{Z}}} 
	g_{mn}^{\alpha  \beta  \gamma  \gamma'}  u^{m} v^{n} . 
\end{align}
The solution with $x=y=0$ corresponds to the solution $g_1$ given in \eqref{eq:g1}, i.e.\ to the unshifted Appell function $F_4$:
\begin{align}
G_1\equiv G_{0,0} ^{\alpha  \beta  \gamma  \gamma'}  (u,v) = 
	\frac{\FF{4}{\alpha,\beta}{\gamma,\gamma'}{u,v}}
		{\Gamma_{1-\alpha} \Gamma_{1-\beta} \Gamma_{\gamma} \Gamma_{\gamma'} } . 
\label{rel:G00F4}		
\end{align}
Hence, $G_1$ inherits its convergence properties from $F_4$, i.e.\ for $x=y=0$ the power series \eqref{eq:SolutionBasisBox} converges if 
\begin{equation}
\label{eq:ConvergenceAppell}
\sqrt{\vert u \vert} + \sqrt{\vert v \vert} < 1.
\end{equation} Here, the sum in \eqref{eq:SolutionBasisBox} effectively only extends over 
$m,n \in \mathbb{N}$ since the above solution \eqref{eqn:Rec-sol-1} implies that
\begin{equation}
g_{m,-n}^{\alpha  \beta  \gamma  \gamma'}  = g_{-m,n}^{\alpha  \beta  \gamma  \gamma'}  = 0 \quad \forall m,n \in \mathbb{N}.
\label{eq:zeros00}
\end{equation}
In fact, this can also be observed directly by inspecting the recurrence relations \eqref{eqn:Rec-1,eqn:Rec-2}. 
Note that if we move $x$ or $y$ slightly away from zero, the sum in 
equation \eqref{eq:SolutionBasisBox} will extend over all of $\mathbb{Z}$
and diverge. 
We assume that a convergent series is only obtained if $x$ and $y$ are chosen 
in such a way that the series terminates at a lower or upper bound for both $m$ and $n$. 
To achieve this we can identify all zeros of the solution \eqref{eqn:Rec-sol-1}, generalizing \eqref{eq:zeros00} for $(x,y)=(0,0)$. This limits us to the following 12 choices for $(x,y)$: 
\begin{align}
&\text{Region I}
&
&\text{Region II}
&
&\text{Region III}
\nonumber
\\\hline
& (0,0) & 
& (-\alpha,0) & 
& (0,-\alpha) \nonumber \\ 
&  (1-\gamma,0)   & 
& (-\beta,0) & 
& (0,-\beta)  \nonumber \\
& (0,1-\gamma')  & 
& (\gamma' - \alpha - 1,1-\gamma') & 
& (1-\gamma,\gamma -\alpha - 1)  \nonumber \\
&(1-\gamma,1-\gamma') & 
& (\gamma' - \beta -1, 1 -\gamma')   & 
&  (1-\gamma,\gamma -\beta - 1)
\label{Values:xy}
\end{align}
Hence, we have 12 solutions of the Yangian PDEs, which are of the form \eqref{eq:SolutionBasisBox} and for which the series terminates. Anticipating their interpretation we have already split these into three categories.

Let us see how this basis of solutions is related to the four functions 
$g_{j=1,2,3,4}$ given in equations \eqref{eq:g1}--\eqref{eq:g4} of the previous \Secref{sec:ParamBox}.
Using the identity \eqref{eq:ReflId2} for Gamma functions, we immediately see that $G_1$ corresponds to the previous solution $g_1$.
For the case $(x,y)= (1 - \gamma, 0)$ we have 
\begin{align}
G_2\equiv G_{1 - \gamma,0} ^{\alpha  \beta  \gamma \gamma'}  (u, v) 
	= u^{1 - \gamma} \sum \limits _{m, n \in \mathbb{Z}} 
	g_{m + 1 - \gamma,n} ^{\alpha  \beta  \gamma \gamma'}  u^{m} v^{n} .
\end{align}
Now, note that 
\begin{align}
g_{m + 1 - \gamma,n} ^{\alpha  \beta  \gamma  \gamma'}  = 
	g_{mn} ^{\alpha + 1 -\gamma , \beta + 1 - \gamma , 2 - \gamma , \gamma'} , 
\label{eq:shift1}	
\end{align}
which follows directly from the properties of the fundamental solution \eqref{eqn:Rec-sol-1}.
We have thus found that 
\begin{align}
\label{eq:SumRelation1}
G_{1 - \gamma,0}  ^{\alpha  \beta  \gamma  \gamma'} (u, v) 
	&= u^{1-\gamma}  G_{0,0} ^{\alpha + 1 -\gamma , \beta + 1 - \gamma , 2 - \gamma , \gamma'} ,
\end{align}
which is related to $g_2$ by a constant factor that can be obtained from equation
\eqref{rel:G00F4}. In a similar fashion, we find the relations 
\begin{align}
\label{eq:SumRelation2}
G_3&\equiv G_{0,1 - \gamma'} ^{\alpha  \beta \gamma  \gamma'}  (u,v) 
 \propto g_3(u,v) ,  \\
G_4&\equiv G_{1 - \gamma,1 - \gamma'} ^{\alpha  \beta  \gamma \gamma'}  (u,v) 	
 \propto g_4(u,v)	. 
\label{eq:SumRelation3}
\end{align}
Modulo overall constants we have thus established the correspondence $G_j \leftrightarrow g_j$ for $j=1,2,3,4$, i.e.\ we have identified the solutions in Region I with the four solutions discussed in \Secref{sec:ParamBox}.

\begin{figure}[t]
\begin{center}
\includegraphicsbox[scale=.65]{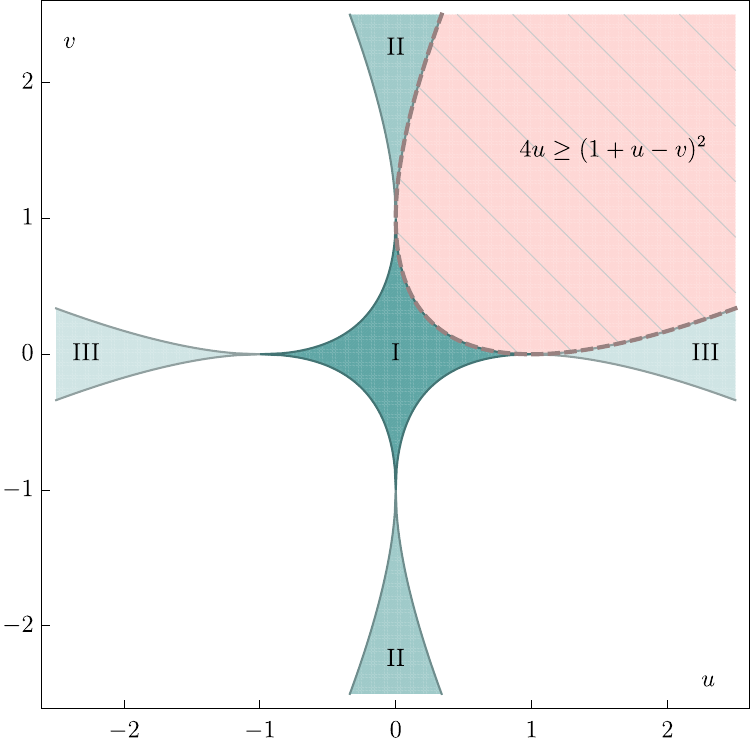}
\end{center}
\caption{
Regions I--III (green) as defined in \protect\eqref{eq:regionI,eq:regionII,eq:regionIII}. The striped (red) area indicates the region of Euclidean physical kinematics, while its complement in the above graph corresponds to Minkowski signature. The dashed boundary between the two regions is given by the line 
$4u = (1+u-v)^2$ or $z=\bar z$, respectively.}
\label{fig:ParameterSpace}
\end{figure}

So what is the meaning of the remaining eight solutions?
For the case $(x,y) = (0,-\alpha)$, we note that 
\begin{equation}
g_{m, n-\alpha} ^{\alpha  \beta  \gamma  \gamma'} = 
	g_{m,-m-n} ^{\alpha ,1+\alpha -\gamma' ,\gamma, 1+\alpha -\beta}, 
	\label{eq:OneShiftg}
\end{equation}
which implies
\begin{align}
G_{0,-\alpha} ^{\alpha , \beta , \gamma , \gamma'} (u,v) 
&= v^{-\alpha} G_{0,0} ^{\alpha ,1+\alpha -\gamma' ,\gamma, 1+\alpha -\beta}  
	\brk*{\sfrac{u}{v} , \sfrac{1}{v} } . 
\end{align}  
Comparing with the convergence condition \eqref{eq:ConvergenceAppell}, we see that in this case the series expansion is hence convergent if 
$ \sqrt{\vert u/v \vert} + \sqrt{\vert 1/v \vert}  < 1$.
Note that we have not found a new solution beyond the four we already encountered. The additional solutions correspond to analytic continuations of the four original series to different regions of kinematical space, see \Figref{fig:ParameterSpace}:
\begin{align}
&\text{Region I:}
&
& \sqrt{\vert u \vert} + \sqrt{\vert v \vert} < 1,
\label{eq:regionI}
\\
&\text{Region II:}
&
& \sqrt{\vert u/v \vert} + \sqrt{\vert 1/v \vert}  < 1,
\label{eq:regionII}
\\
&\text{Region III:}
&
& \sqrt{\vert v/u \vert} + \sqrt{\vert 1/u \vert}  < 1.
\label{eq:regionIII}
\end{align}

In order to see the relation to the original solutions explicitly, we can employ the functional identity \eqref{eq:FuncIdF4} for the Appell function $F_4$ given in \Appref{sec:Identities}. This yields the relation 
\begin{align}
 G_{0,-\alpha} ^{\alpha  \beta  \gamma \gamma'} (u,v) = 
 &	e^{-i \pi  \alpha }\frac{ \Gamma_{1-\gamma'} \Gamma_{\gamma'}}
	{\Gamma_{\gamma'-\alpha } \Gamma _{1+\alpha-\gamma'}}
   G_1\nonumber \\
+&e^{-i \pi (1+\alpha -\gamma')}\frac{ \Gamma _{2-\gamma'} \Gamma _{\gamma'-1}}
   {\Gamma _{1-\alpha } \Gamma _{\alpha }}
    G_3. 
\label{FuncRel1}   
\end{align}
Similar relations can be established for the remaining choices of $(x,y)$ 
as well. We have thus derived from scratch that the solution to the Yangian PDEs is described by a linear combination of four series converging around $u=v=0$:
\begin{align}
&\phi = c_1 G_1 + c_2 G_2
	+ c_3 G_3+ c_4 G_4 . 
\label{Phi-Ansatz}	
\end{align}
Here, we have suppressed the dependence on the cross ratios $u,v$ as well 
as the parameters $\alpha ,\beta ,\gamma ,\gamma'$ for the functions $G_j$ and the 
coefficients $c_j$. 

\paragraph{Permutation Symmetries.}

As anticipated in \Secref{sec:ParamBox},
the coefficients can be constrained by employing the invariance of the integral 
$I_4$ under simultaneous permutations of the external points $x_j$ and
the associated propagator powers $a,b,c,d$ entering the solutions through the relations \eqref{eq:RelsGreekLatin}.
In order to derive the consequences of permutation invariance in a compact
form, we consider the generators $\sigma_1 ^x = (1234)$ and $\sigma_2 ^x= (12)$ of 
the permutation group $S_4$, which act on the external legs of the Feynman diagram. These generators act on the cross ratios and parameters 
$\alpha ,\beta ,\gamma ,\gamma'$ as 
\begin{align*}
\sigma_1:& 
\begin{cases}
	(u,v) \mapsto (v,u), \\
	(\alpha, \beta, \gamma, \gamma') \mapsto 
		(1+\beta-\gamma, 1+\alpha-\gamma, \gamma', 2-\gamma) ,
\end{cases}  \\
\sigma_2:& 
\begin{cases}
	(u,v) \mapsto (u/v,1/v), \\
	(\alpha, \beta, \gamma, \gamma') \mapsto 
		(1+\beta-\gamma', \beta, \gamma, 1+\beta-\alpha) .
\end{cases} 
\end{align*}
We recall the relation between the function $\phi(u,v)$ and the integral
\begin{align}
I_4 = x_{14}^{2\gamma'-2}  x_{34}^{2\gamma -2}  
		x_{13}^{-2\beta} x_{24}^{-2\alpha} 
	\phi ^{\alpha\beta \gamma \gamma'}  (u,v) ,  
\end{align}
and note the functional relations that follow from invariance under $\sigma_1$ and 
$\sigma_2$, respectively:
\begin{align}
\phi ^{\alpha\beta\gamma\gamma'} (u,v) &= 
	u^{1-\gamma} \phi ^{1+\beta-\gamma, 1+\alpha-\gamma, \gamma', 2-\gamma} (v,u) 
	\label{perm1} , \\
\phi ^{\alpha\beta \gamma\gamma'} (u,v) &= 
	v^{-\beta} \phi ^{1+\beta-\gamma', \beta, \gamma, 1+\beta-\alpha} \brk*{\sfrac{u}{v},\sfrac{1}{v}} . 	
\label{perm2}	
\end{align}
The invariance under $\sigma_1$ allows to express the coefficients 
$c_2, c_3$ and $c_4$ in the ansatz \eqref{Phi-Ansatz} in terms of $c_1$:
\begin{align}
c_2 &= c_1 \circ \sigma_1 , & 
c_4 &= c_1 \circ \sigma_1^2, & 
c_3 &= c_1 \circ \sigma_1^3 . 
\end{align} 
The additional invariance under $\sigma_2$ implies functional relations 
for $c_1$ as a function of the parameters $\alpha ,\beta ,\gamma ,\gamma'$. 
The simplest way to state these relations is to note that the function
\begin{align}
N_4 ^{\alpha \beta \gamma \gamma'} = \pi^{-4}
	\sin \pi \alpha \sin \pi \beta \sin \pi \gamma \sin \pi \gamma' 
	\,c_1 ^{ \alpha \beta\gamma \gamma'} 
\end{align}
is invariant under both the actions of $\sigma_1$ and $\sigma_2$ on the 
parameters. As a function of the parameters $a,b,c,d$, $N_4$ is hence invariant 
under all permutations of its arguments. This allows us to express all coefficients 
appearing in our ansatz \eqref{Phi-Ansatz} in terms of the coefficient 
$N_4 ^{\alpha \beta \gamma \gamma'}$. 


\paragraph{Overall Constant.}
The above requirement does not determine the coefficient 
$N_4 ^{\alpha \beta \gamma \gamma'}$ uniquely and we employ the coincidence 
limit $2\to 1$ of external points of the Feynman diagram in order to fix it. Applying this limit to the box integral, 
we find
\begin{equation}
\lim_{2\to 1} I_4 = \int 
	\frac{\dd^D{x_0}}{x_{10}^{2(a+b)}x_{30}^{2c}x_{40}^{2d}} 
	=\frac{\STconst_{a+b,c,d}}{x_{13}^{2d'}x_{34}^{2(a+b)'}x_{14}^{2c'}}.
\end{equation}
On the other hand, for the cross ratios the limit $2\to 1$ implies that
$(u,v) \to (0,1)$
and we can write
\begin{equation}
\label{eq:CoincLimitBox}
\lim_{2\to 1} I_4 =
	\frac{\lim_{(u,v)\to (0, 1)}\phi(u,v)}
		{x_{13}^{2d'}x_{34}^{2(a+b)'}x_{14}^{2c'}}.
\end{equation}
We thus read off that
\begin{equation}
\lim_{(u,v)\to (0, 1)} \phi(u,v)
=
\STconst_{a+b,c,d},
\end{equation}
On the basis functions $G_j$ appearing in our ansatz \eqref{Phi-Ansatz}, 
this limit acts as (we assume $\gamma < 1$)
\begin{align*}
G_0 &\to \frac{ \Gamma_{\gamma'-\alpha-\beta}}
	{\Gamma _{1-\alpha } \Gamma _{1-\beta}
		\Gamma _{\gamma} \Gamma _{\gamma'-\alpha} \Gamma _{\gamma'-\beta}} , & 
G_1&\to	0 , & \\	
G_2 &\to  \frac{ \Gamma_{\gamma'-\alpha-\beta}}
	{\Gamma _{1-\alpha } \Gamma _{1-\beta} 
		\Gamma _{\gamma} \Gamma _{\gamma'-\alpha} \Gamma _{\gamma'-\beta}} , & 	
G_3 &\to	0.		
\end{align*}
Here we have employed the identity \eqref{eq:GaussAt1}
for the Gau{\ss} hypergeometric function to end up with expressions in terms of Gamma functions.
Hence, we have obtained 
the relation 
\begin{align*}
\lim_{(u,v)\to (0, 1)}\phi(u,v) &
= 
	N_4^{\alpha \beta \gamma\gamma'} \Gamma_{1-\gamma}
		 \Gamma_{\gamma'} \Gamma _{1-\gamma'} 
		 \Gamma_{\gamma' - \alpha - \beta}  \\
& \times  \brk*{ \frac{\Gamma_{\alpha} \Gamma_{\beta}}
				{\Gamma_{\gamma' - \alpha} \Gamma_{\gamma' - \beta} } 
		- \frac{\Gamma_{1+\alpha-\gamma'} \Gamma_{1+\beta-\gamma'}}
				{\Gamma_{1 - \alpha} \Gamma_{1 - \beta} }  } \\
&= 	 \frac{\pi ^{D/2} \Gamma_{1-\gamma} \Gamma_{1+\alpha-\gamma'} \Gamma_{\beta}}
		{\Gamma_{1+\alpha+\beta-\gamma'}\Gamma_{1+\beta-\gamma}
			\Gamma_{2+\alpha-\gamma-\gamma'}} ,  	 		
\end{align*}
which we solve for $N_4^{\alpha \beta\gamma\gamma'}$ to find
\begin{align*}
N_4^{\alpha \beta \gamma \gamma'} = 
	\frac{\pi ^{2+\alpha+\beta-\gamma-\gamma'} }
		{\Gamma_{\alpha} \Gamma_{1+\beta-\gamma} 
			\Gamma_{1+\beta-\gamma'} \Gamma_{2+\alpha - \gamma - \gamma'}}
	= \frac{\pi ^{D/2} }
		{\Gamma_{a} \Gamma_{b} \Gamma_{c} \Gamma_{d}} . 		
\end{align*}
Note that the latter form makes the permutation symmetry manifest. 
We have thus bootstrapped the $D$-dimensional box integral \eqref{eq:BoxIntD}
with generic propagator powers and obtained the result 
\begin{align}
\phi = 
	N_4^{\alpha \beta \gamma \gamma'}
	\brk!{\tilde G_1-\tilde G_2-\tilde G_3+\tilde G_4},
\end{align}
where for $j=1,\dots,4$ we have defined 
\begin{align}
\tilde G_j
= \frac{\pi ^4 \csc \pi \gamma \csc \pi \gamma' G _{x_j,y_j} ^{\alpha, \beta, \gamma, \gamma'} }
	{\sin \pi (\alpha + x_j + y_j) \sin \pi (\beta + x_j + y_j) } .
\end{align}
Here $(x_j,y_j)$ label the four shifts in Region I of \eqref{Values:xy} and
we remind the reader that the propagator powers $a,b,c,d$ are related to the greek parameters via \eqref{eq:RelsGreekLatin}. 

It may be useful to summarize the algorithmic steps that allowed 
us to bootstrap the above box integral:
\smallskip

\begin{minipage}{0.45\textwidth}
\begin{enumerate}
\item translate the Yangian PDEs into recurrence equations \eqref{eqn:Rec-1,eqn:Rec-2},
\item find a fundamental solution \eqref{eqn:Rec-sol-1},
\item find all zeros \eqref{Values:xy} of the fundamental solution,
\item classify the zeros by their kinematic region \eqref{Values:xy}, 
\item in a given kinematic region, use the permutation symmetries and coincidence limits to fix the linear combination.
\end{enumerate}
\end{minipage}
\medskip

\noindent Importantly, we note that as external input we have used the convergence properties of the Appell function $F_4$
as given in the literature. Moreover, classification of the kinematic regions (point 4.) can be achieved through investigation of shift identities of the form \eqref{eq:OneShiftg}; similar identities are not guaranteed to exist for fundamental solutions \`a la \eqref{eqn:Rec-sol-1} for different integrals.
\section{Six-Point Double Box}
\label{sec:DoubleBox}

Being a member of the class of fishnet Feynman graphs discussed in \cite{Chicherin:2017cns,Chicherin:2017frs}, also the double box integral is invariant under the conformal Yangian algebra:
 \begin{equation}
\hat I_{3,3}= \int  \frac{\dd^4 x_0 \dd^4 x_{0'}}{x_{10}^{2}x_{20}^{2}x_{30}^{2} x_{00'}^{2}x_{40'}^{2}x_{50'}^{2}x_{60'}^{2}}
=
\includegraphicsbox[scale=.8]{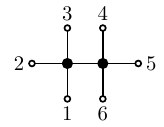}
 \end{equation}
 In this case conformal symmetry dictates that $I_{3,3}$ is of the form
 \begin{equation}
 \hat I_{3,3}=\frac{1}{x_{13}^{2}x_{25}^{2}x_{46}^{2}} \hat\phi_{3,3}(u_1,\dots, u_9),
 \end{equation}
 with a conformally invariant function $\hat \phi_{3,3}$ of nine cross ratios that we
 define as in \Appref{sec:CrossRatios}, cf.\ \cite{Nandan:2013ip}.
The partial differential equations arising from the Yangian level-one
symmetry read
\begin{equation}
\PDE_{jk} \,\hat\phi_{3,3}(u_1,\dots,u_9)=0,
\end{equation}
with 6 of the 15 differential operators $\PDE_{jk}$ given by
\begin{allowdisplaybreaks}
\begin{widetext}
\begin{align}
\PDE_{12}=
& -\theta_6 ^2
	+u_6  \left(D_{1 6 8}+1\right) D_{3 6 5}
	+u_5 u_6 \left(D_{1 6 8} + 1 \right) \left(D_{2 5 3 4} + 1 \right)
	- u_6 u_8 D_{3 6 5} D_{1 9 2 8} 
	+ u_4 u_5 u_6 D_{1 4 2} \left(D_{1 6 8}+1\right)
	\nonumber \\
& +u_6 u_8 u_9 D_{3 6 5} D_{3 9 2}
	-u_5 u_6 u_7 u_8 D_{1 9 2 8} \left(D_{2 5 3 4}+1\right), 
	 \nonumber \\
\PDE_{13}=& \theta _8 \left(D_{1 6 8} + 1 \right)
	-u_8 D_{1 9 2 8} D_{5 8 6 7}	
	-u_7 u_8 D_{4 7 5} D_{1 9 2 8}
	+u_8 u_9 D_{3 9 2} D_{5 8 6 7}	,
	 \nonumber\\
\PDE_{14}=&\left(\theta _1-\theta _9\right) D_{1 9 2 8}
	-u_1 D_{1 4 2} \left(D_{1 6 8}+1\right)
	+u_9 D_{3 9 2} \left(D_{7 9 8}+1\right), 
	 \nonumber\\
\PDE_{15}=& -\theta _2 D_{3 9 2}
	+ u_2 D_{1 9 2 8} \left(D_{2 5 3 4} + 1 \right)	
	-u_1 u_2 \left(D_{1 6 8} + 1 \right) \left(D_{2 5 3 4} + 1 \right)	
	+u_2 u_4 D_{4 7 5} D_{1 9 2 8}, 
	 \nonumber\\
\PDE_{16}=& \left(\theta _3-1\right) \theta _3
	-u_3 D_{3 6 5} D_{3 9 2}
	+u_2 u_3 D_{3 6 5} D_{1 9 2 8}
	-u_3 u_5 D_{3 9 2} D_{5 8 6 7}
	 -u_1 u_2 u_3 \left(D_{1 6 8}+1\right) D_{3 6 5}
	 \nonumber \\	
& 	-u_3 u_5 u_7 D_{3 9 2} \left(D_{7 9 8} + 1 \right)
	+u_2 u_3 u_4 u_5 D_{1 9 2 8} D_{5 8 6 7}, 
	 \nonumber\\
\PDE_{25}=& \left(\theta _3-\theta _4\right) \left(D_{2 5 3 4}+1\right)
	+u_3 D_{3 6 5} D_{3 9 2}
	-u_4 D_{1 4 2} D_{4 7 5}.
	\label{eq:PDEsDoubleBox}
\end{align}
\end{widetext}
\end{allowdisplaybreaks}
Above, for compactness, the Euler operators $\theta_j=u_j\partial_{u_j}$ are packaged into 
\begin{align}
D_{ijk}&=\theta_i+\theta_j-\theta_k,
\nonumber\\
D_{ijkl}&=\theta_i+\theta_j-\theta_k-\theta_l.
\end{align}
Moreover, the remaining 9 Yangian differential operators $\PDE_{jk}$ of the set \eqref{eq:PDEeq0}, which annihilate $\phi_{3,3}$, are obtained from the following permutations of cross ratio labels:
\begin{align}
\sigma^u_3 &= (16)(25)(34)(79) , & 
\sigma^u_4 &= (19)(34)(67) .
\label{eq:DBPerms}
\end{align}
Notably, these permutations of the cross ratios $u_i$ leave $\phi_{3,3}$ 
invariant:
\begin{align}
&\phi_{3,3}(u_6,u_5,u_4,u_3,u_2,u_1,u_9,u_8,u_7)
\nonumber\\
&\qquad=\phi_{3,3}(u_1,u_2,u_3,u_4,u_5,u_6,u_7,u_8,u_9),
\label{eq:FuncId1}
\\
&\phi_{3,3}(u_6,u_5,u_4,u_3,u_2,u_1,u_9,u_8,u_7)
\nonumber\\
&\qquad= \phi_{3,3}(u_1,u_2,u_3,u_4,u_5,u_6,u_7,u_8,u_9).
\label{eq:FuncId2}
\end{align}
These identities result from imposing invariance of the double box 
Feynman graph under the permutations
\begin{align}
\sigma^x_3 &= (14)(25)(36) , & 
\sigma^x_4 &= (13)(46) ,
\end{align}
of the six external legs, respectively.
An important point to note is that these permutations not only leave the integral invariant, but also the level-one momentum generator \eqref{eq:Phat}. Therefore, the full invariance equation \eqref{eq:PhatonIn} stays invariant under this permutation, which makes it easy to identify pairs of differential equations that are related by the corresponding functional identity.
Further functional identities generalizing the invariance under the permutations 
above for the box integral, are listed in \Appref{sec:FuncIdentitiesDoubleBox}.

Similar differential equations as given in \eqref{eq:PDEsDoubleBox} can be written down for the double box integral with generic propagator powers 
\begin{equation}
\label{eq:DefDoubleBox}
I_{3,3}=\int  \frac{\dd^D x_0 \dd^D x_{0'} }{x_{10}^{2a} x_{20}^{2b} x_{30}^{2c} x_{00'}^{2\ell} x_{40'}^{2d} x_{50'}^{2e} x_{60'}^{2f}} = V_{3,3} \, \phi_{3,3} \, ,
\end{equation}
where we have stripped off a prefactor
\begin{align}
V_{3,3} &= x_{13}^{2\ell-D} x_{14}^{D-2\ell} x_{15}^{-2d-2e} x_{16}^{2d+2e-2a} x_{26}^{-2b} x_{36}^{D-2c-2\ell} \nonumber\\
&\hspace{14pt}  x_{46}^{2\ell-2d-D} x_{56}^{2d} \,,
\end{align}
and conformality requires $a+b+c+\ell=D$ and $d+e+f+\ell=D$.

Solving these differential equations in nine variables is obviously  a much more involved task than for the two-variable box function. Moreover, the double box integral has less permutation symmetries than the totally symmetric cross integral. It is thus reasonable to approach this problem from a more symmetric direction and to consider a simpler situation.


\section{Hexagon}
\label{sec:Hexagon}

The double box integral in $D$ dimensions is related to the $(D+2)$-dimensional hexagon 
 via the following simple differential equation relating the respective conformally invariant functions \cite{Paulos:2012nu}
\begin{equation}
\partial_{u_8} \phi_{3,3}(u_1,\dots,u_9,D)
=-\frac{\pi^{D/2-1}}{\Gamma_{\ell}}
\phi_6(u_1,\dots,u_9,D+2),
\label{eqn:PSVdiffeq}
\end{equation}
which holds true for $D/2-\ell=1$ and with $\phi_{3,3}$ and $\phi_6$ as defined in \Appref{sec:feynpara}. 
In fact, using the expressions provided in \Appref{sec:feynpara} it can be shown that the following slightly stronger equation holds true
\begin{equation}
\label{eq:DoubleBoxHexagon}
\includegraphicsbox[scale=1]{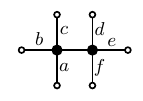}
=\frac{\pi^{D/2-1}}{\Gamma_{\ell}} \int_{u_8}^\infty \dd u' \, 
\includegraphicsbox[scale=1]{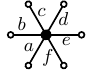}.
\end{equation}
Note that here the Feynman diagrams do not represent the full integrals but rather the conformally invariant functions $\phi_{3,3}(u_1,\dots,u_9)$ on the left hand side and $\phi_6(u_1,\dots,u_7,u',u_9)$ on the right hand side, respectively.
The above relation implies that the hexagon integral obeys similar differential equations as the double box.
In fact, we can give an argument independent of the double box, which shows that the hexagon is Yangian invariant in three and six spacetime dimensions. Firstly, in three dimensions, the hexagon is simply the fundamental Yangian-invariant vertex, similar to the box integral in four dimensions \cite{Chicherin:2017frs}. In six dimensions, Yangian-invariance follows from the following two observations: i)~In \cite{Chicherin:2017frs} it was noted that six-dimensional Feynman graphs with propagator weights 2 and built from three-point vertices, are Yangian invariant (similar results hold for deformed propagator powers), ii)~Using the star-triangle relation \eqref{eq:StarTriangle}, the hexagon multiplied by external propagators on the left hand side can be related to a three-point graph shown on the right hand side:
\begin{equation}\label{eq:hexis3pt}
\includegraphicsbox[scale=1]{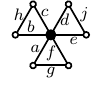}
= 
\STconst^{-1}_{a'f'g'}
\STconst^{-1}_{b'h'c'}
\STconst^{-1}_{d'j'e'}
\includegraphicsbox[scale=1]{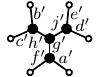}
\end{equation}
Here the star-triangle relation requires the constraints $a+f+g=D/2$, $b+h+c=D/2$ and $d+j+e=D/2$.

Let us now consider the resulting Yangian constraints for the conformal hexagon integral in the form
\begin{equation}
\label{eq:HexagonDefConformal}
\includegraphicsbox[scale=1]{FigHexagonPropWeights.pdf}
= \! \int  \frac{\dd^D x_0 }{x_{10}^{2a}x_{20}^{2b}x_{30}^{2c} x_{40}^{2d}x_{50}^{2e}x_{60}^{2f}}
= V_6 \,\phi_6(w_1,\dots,w_9).
\end{equation}
Here we have $a+b+c+d+e+f=D$ and  the prefactor
\begin{equation}
\label{eq:HexagonPre}
V_6=x_{16}^{D-2a-2f}x_{26}^{-2b} x_{36}^{-2c}x_{46}^{-2d}x_{56}^{-2e-2f+D}x_{15}^{2f-D}.
\end{equation}
Moreover, we have redefined the cross ratios employed in the conformal 
parametrization above according to
\begin{align}
w_1&=u_3,
&
w_2&=u_3 u_5,
&
w_3&=u_3u_5u_7,
\nonumber\\
w_4&=u_9,
&
w_5&=u_2u_3u_9,
&
w_6&=u_2u_3u_4u_5u_9,
\nonumber\\
w_7&=u_8u_9,
&
w_8&=u_1u_2u_3u_8u_9,
&
w_9&=u_6u_8u_9.
\label{eq:alt9crosscr}
\end{align}
These turn out to be convenient in order to write the fundamental solution to 
the Yangian recurrence equations in the  form of a Taylor series. 

Having established the Yangian symmetry and the conformal parametrization 
of the hexagon integral, we employ the evaluation parameters given in 
equation \eqref{eq:HexEvalParam} 
and apply the Yangian level-one generator $\gen{\widehat P}^{\mu}$ to the 
above expression. This yields an invariance equation of the form 
\eqref{eq:PhatonIn}, from which we read off the 15 partial differential 
equations collected in \Appref{sec:hexPDEs}. We then employ a 
series ansatz in terms of the cross ratios \eqref{eq:alt9crosscr}:
\begin{align}
\label{eq:SeriesAnsatz}
\phi_6(w_j) = \sum_{n_1,\dots,n_9} h_{n_1 \dots n_9} \,w_1^{n_1}\dots  w_9^{n_9} . 
\end{align}
Here, for convenience we have set 
\begin{equation}
\label{eq:h1func}
h_{n_1 \dots n_9}
 = f_{n_1 \dots n_9}\prod\nolimits_{j=1}^9 \Gamma_{n_j+1}^{-1} .  
\end{equation}
The recurrence equations for $f_{n_1\dots n_9}$, which follow from imposing the Yangian PDEs on the above series ansatz, are listed 
in \Appref{sec:RecEqsHex}. Notably, these equations appear too complicated to be solved by elementary means. However, a fundamental solution to these recurrences
can be obtained from the Feynman parameter representation of the hexagon integral given in \eqref{eqn:PhisFeynPara}. 
Taylor-expanding this representation in the cross ratios and integrating order by order yields the expression
\begin{align}
f_{n_1 \dots n_9}
 &= \frac{1}{
\Gamma_{1-M_1}
\Gamma_{1-M_2}
\Gamma_{1-M_3}
\Gamma_{1-M_4}
\Gamma_{M_5}
\Gamma_{M_6} },
\label{eq:FundSol}
\end{align}
where we use the shorthands
\begin{align}
 M_1&=\alpha+{\scriptstyle\sum_{k=1}^9} n_k,
 \label{eq:MHexa}
 \\
 M_2&=\beta_1+n_1+n_5+n_8+n_9,
  \nonumber
 \\
 M_3&= \beta_2+n_2+n_6+n_7+n_8,
  \nonumber
  \\
 M_4&= \beta_3+n_3+n_4+n_5+n_6,
  \nonumber
  \\
 M_5&=\gamma_1+n_1+n_2+n_3+n_5+n_6+n_8,
  \nonumber
  \\
 M_6&=\gamma_2+n_4+n_5+n_6+n_7+n_8+n_9 , \nonumber
 \end{align}
and the greek parameters encode the propagator powers of \eqref{eq:HexagonDefConformal}:
\begin{align}
&\alpha=\frac{D}{2}-f, \quad \beta_1=b, \quad \beta_2= c, \quad\beta_3= d,  \\
&\gamma_1=1+\frac{D}{2}-a-f,\hspace{15pt} \gamma_2=1+\frac{D}{2}-e-f . \nonumber
\end{align}
The above function $h_{n_1\dots n_9}$ defined through \eqref{eq:h1func} represents an analogue of the fundamental solution \eqref{eqn:Rec-sol-1} for the box integral.
Plugging this solution to the recurrence equations into the series \eqref{eq:SeriesAnsatz}, we obtain a Yangian invariant that can be identified with a (Srivastava--Daoust) Lauricella function \cite{srivastava1969certain,srivastava1985multiple}
\begin{equation}
H_1 = \sum_{n_1,\dots,n_9 \in \mathbb{Z}} 
	h_{n_1,\dots,n_9} w_1^{n_1}\dots  w_9^{n_9}.
\end{equation}
This is the analogue of the function 
$G_1$ given in \eqref{rel:G00F4} in the bootstrap of the box integral.  
Similar to the case of the box, $H_1$ yields the analytic part of 
the hexagon integral \eqref{eq:HexagonDefConformal} in the conformal variables $w_j$:
\begin{equation}
\includegraphicsbox[scale=1]{FigHexagonPropWeights.pdf}
= c_1 H_1 +\text{non-analytic}.
\end{equation}
Here the coefficient $c_1$ is given by 
\begin{equation}
c_1 =\frac{\pi^{2+\alpha +\beta_1+\beta_2+\beta_3-\gamma_1-\gamma_2} \Gamma_{1-\beta_1} \Gamma_{1-\beta_2} \Gamma_{1-\beta_3} \rho_\alpha \rho_{\gamma_1} \rho_{\gamma_2}}{\Gamma_{1+\alpha -\gamma_1} \Gamma_{1+\alpha -\gamma_2} \Gamma_{2+\beta_1+\beta_2+\beta_3-\gamma_1-\gamma_2}}, 
\end{equation}
with the shorthand $\rho_x = \Gamma_{x} \Gamma_{1-x}$.
As before, additional solutions of the Yangian invariance conditions can be 
obtained by summing over a shifted lattice with base point 
$(x_1, \ldots, x_9)$:
\begin{equation}
\label{eq:SolutionBasisHexaon}
H_{x_1\dots x_9} = 
	\sum_{n_j \in x_j + \mathbb{Z}}
	h_{n_1\dots n_9} \,w_1^{n_1}\dots  w_9^{n_9}.
\end{equation}
Obviously we have $H_1=H_{0\dots0}$.
 Restricting to base points for which the series terminates 
in all nine parameters, we find 2530 possible sets $(x_1,\dots, x_9)$, which
compares to the situation of the box integral as follows:
\begin{equation}
\begin{tabular}{l|p{1.3cm}|p{1.3cm}}
&Box & Hexagon
\\\hline
Variables&2&9
\\
Series &12 &2530
\end{tabular}
\end{equation}
As for the case of the box integral, we expect that these Yangian invariants are series representations converging in different domains, but are linked 
by functional relations similar to equation \eqref{eq:FuncIdF4}. One may thus expect that the total 
number of Yangian invariants is lower than the number 2530 of series 
representations found above. 

In the algorithm outlined at the end of \Secref{sec:recsBox}, the next step 
would be to classify the above sets $(x_1,\dots, x_9)$, i.e.\ the zeros of the fundamental solution $h_{n_1\dots n_9}$, by their kinematic region. For the box integral this can most efficiently be done by employing the
shift identities listed in \Appref{sec:shifts}. However, it is not clear that similar shift identities
exist for the given fundamental solution of the hexagon. This obscures the identification
of a linear combination of the above series that represents the full hexagon integral.
Moreover, a full analysis of the domain of convergence of all series representations 
seems to require a significant improvement in the current understanding of the properties of the above generalized Lauricella functions. We thus leave further steps into these directions for future work.

As argued above, the double box integral is even more involved than the hexagon considered in this section. This makes it clear that gaining full control over the hexagon bootstrap is a natural prerequisite for further investigations of the double box discussed in \Secref{sec:DoubleBox}.

\section{Recursive Structure and Overall Constants}
\label{sec:RecStruc}


As demonstrated in the previous \Secref{sec:Hexagon}, Yangian symmetry does not fix the considered six-point integrals
completely. This underlines the need for further constraints required to eventually bootstrap these integrals. 
As argued in \Secref{sec:recsBox} for the case of the box integral, also the
considered six-point integrals can be recursively related to the star-triangle
relation which can thus be used to e.g.\ fix their overall constants:
\begin{equation}
\includegraphicsbox[scale=.9]{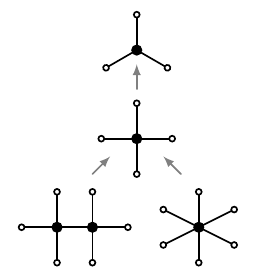}
\end{equation}
While the four-point situation was already discussed in \Secref{sec:recsBox} (see \eqref{eq:CoincLimitBox}), let us explain the six-point cases in more detail.


\paragraph{Hexagon.}
In the case of the hexagon we can take a coincidence limit for three external points, e.g.\ $2\to 1$, $3\to1$, $4\to 1$, to obtain the triangle integral of \eqref{eq:StarTriangle}:
\begin{equation}
\label{eq:HexagonLimitIntegral}
\lim_\limitstack{ 2\to 1}{3\to1}{4 \to 1} \,I_6=
\int  \frac{\dd^D x_0 }{x_{10}^{2(D-e-f)}x_{50}^{2e} x_{60}^{2f}}
=
 \frac{\STconst_{D-e-f,e,f}}{x_{15}^{2f'}x_{56}^{-2(e+f)'}x_{16}^{2e'}}.
\end{equation}
Note that taking only two of the above coincidence limits yields the box integral at an intermediate step.
On the other hand, for the above cross ratios \eqref{eq:alt9crosscr} the triple coincidence limit implies 
\begin{equation}
w_j \to \hat w_j, 
\qquad \text{with}\quad \hat w_{j=1,2,3}= 1, 
\quad \hat w_{j>3}= 0,
\end{equation}
and we can evaluate the limit on the right hand side of \eqref{eq:HexagonDefConformal} to find
\begin{equation}
\lim_\limitstack{ 2\to 1}{3\to1}{4 \to 1} \,I_6
 \frac{\lim_{w_j\to\hat w_j } \phi_6(w_1,\dots,w_9)}{x_{15}^{2f'}x_{56}^{-2(e+f)'}x_{16}^{2e'}}.
\end{equation} 
Comparing to \eqref{eq:HexagonLimitIntegral} we thus read off that
\begin{equation}
\lim_{w_j\to\hat w_j} \phi_6(w_1,\dots,w_9)
=
\STconst_{a+b,c+d,e+f}.
\end{equation}
Note that we can similarly take coincidence limits of different external points leading to further equations which constrain the coefficients of Yangian invariant functions and in particular the overall constant of the integral.


\paragraph{Double Box.}

The case of the double box integral is slightly more involved.
Consider the conformal double box with parameters obeying $a+b+c+\ell=D$ and $\ell+e+f+g=D$:
\begin{equation}
I_{3,3}=\int  \frac{\dd^D x_0 \dd^D x_{0'} }{x_{10}^{2a} x_{20}^{2b} x_{30}^{2c} x_{00'}^{2\ell} x_{40'}^{2d} x_{50'}^{2e} x_{60'}^{2f}}.
\end{equation}
We now take the coincidence limit $2\to 1$ and $5\to4$ of the external points 
such that
\begin{equation}
\lim_{\genfrac{}{}{0pt}{2}{5\to 4}{2\to 1}} I_{3,3}
=\int  \frac{\dd^D x_0 \dd^D x_{0'} }{x_{10}^{2(a+b)} x_{30}^{2c} x_{00'}^{2\ell} x_{40'}^{2(d+e)} x_{50'}^{2f}},
\end{equation}
and we use the star-triangle relation \eqref{eq:StarTriangle}
on the first integral to find the box integral
\begin{equation}
\lim_{\genfrac{}{}{0pt}{2}{5\to 4}{2\to 1}} I_{3,3}
=\frac{\STconst_{a+b,c,\ell}}{x_{13}^{2\ell'}}\int  \frac{ \dd^D x_{0'}\,}{x_{10'}^{2c'} x_{30'}^{2(a+b)'} x_{40'}^{2(d+e)} x_{60'}^{2f}}.
\end{equation}
Note that the sum of propagators in the remaining box integral gives $\ell+d+e+f=D$, i.e.\ the integral has conformal Yangian symmetry. 
We can take a further coincidence limit $4\to 3$ and use again the star-triangle relation to find
\begin{align}
\lim_\limitstack{5\to4}{2\to1}{4\to3} \,I_{3,3}
&=
\frac{\STconst_{a+b,c,\ell}}{x_{13}^{2\ell'}}
\int \frac{\dd x_{0'}}{x_{10'}^{2c'}x_{30'}^{2(a+b)'+2(d+e)}x_{60'}^{2f}}
\nonumber\\
&=
\frac{\STconst_{a+b,c,\ell}\STconst_{D/2-c,D/2+c-f,f}}{x_{13}^{2(D-\ell-f)}x_{16}^{2(f-c)} x_{36}^{2c} }
\end{align}
For the cross ratios the above consecutive triple coincidence limit corresponds to
\begin{equation}
w_j \to \hat w_j, 
\qquad \text{with}\quad \hat w_{j=1,4,5,7,8}= 1, 
\quad \hat w_{j=2,3,6,9}= 0.
\end{equation}
Hence, the limit on right hand side of \eqref{eq:DefDoubleBox} can be written as
\begin{equation}
\lim_\limitstack{5\to4}{2\to1}{4\to3} \,I_{3,3}
=
 \frac{\lim_{w_j\to \hat w_j} \phi_{3,3}(w_1,\dots,w_9)}{x_{13}^{2(D-\ell-f)}x_{16}^{2(f-c)} x_{36}^{2c} }.
\end{equation}
We thus read off that
\begin{align}
\lim_{w_j\to \hat w_j} \phi_{3,3}(w_1,\dots,w_9)
=
\STconst_{a+b,c,\ell}\STconst_{D/2-c,D/2+c-f,f}.
\end{align}
Again, this is only the result of one possible coincidence limit and we can obtain further constraints by taking other limits.

\section{Yangian Invariants and Mellin--Barnes Integrals}
\label{sec:MB}

\begin{figure}[t]
\begin{center}
\includegraphicsbox[scale=.65]{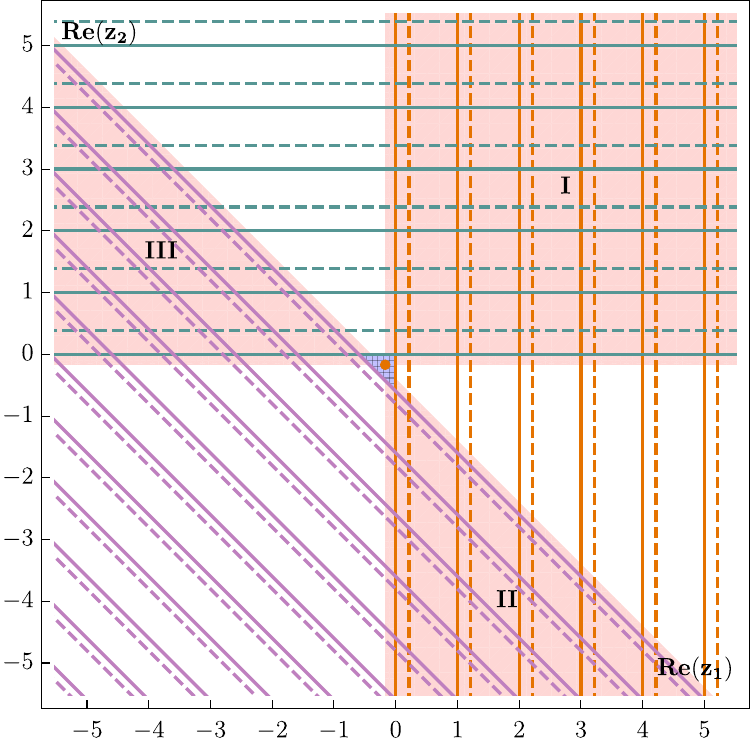}
\end{center}
\caption{Singularity structure of the Mellin--Barnes integrand \protect\eqref{eqn:MBintegrandboxint}. Colored lines correspond to poles of Gamma functions. The orange dot in the center of the plot marks the base point of the integration and can be moved inside the meshed (blue) triangle without changing the value of the integral. Red cones mark the regions in which residues need to be summed to obtain a valid series representation of the integral.}
\label{fig:MBPoles}
\end{figure}
Integrability is very constraining and if properly understood one can expect that it completely fixes physical observables through the underlying symmetry constraints. In \cite{Chicherin:2017cns,Chicherin:2017frs} it was shown that certain Feynman graphs provide a means to obtain an infinite class of Yangian invariants. In the previous section we have shown that these Yangian invariants have a fine structure, i.e.\ there are more elementary Yangian invariants whose linear combination is selected by imposing further symmetries (e.g.\ permutation invariance) of the considered Feynman integrals. So what is the construction principle underlying these more elementary Yangian invariants and what is the most natural way to fix their linear combination? In order to get more insights into this, it is useful to compare the above construction to the Mellin--Barnes technique for obtaining certain Feynman integrals.

Let us discuss the box integral in more detail. Its Mellin--Barnes representation is most conveniently obtained by iteratively applying the rule
\begin{align}
\frac{1}{(A+B)^\lambda}= \frac{1}{\Gamma_{\lambda}}\frac{1}{2 \pi i} \int\limits_C \dd z \frac{A^{z}}{B^{\lambda+z}}\Gamma_{-z} \Gamma_{\lambda+z}  ,
\label{eqn:MBbasicformula}
\end{align}
to the Feynman parameter representation in \Appref{sec:feynpara} and integrating out the Feynman parameter integrals. Here the contour $C$ extends from $-i \infty$ to $+i \infty$ and is chosen such that it separates the two pole series of the Gamma functions. If the intersection between the areas characterized by $\text{Re}(-z)>0$ and $\text{Re}(\lambda+z)>0$ is non-empty, the contour $C$ can be taken to be a straight line $c+i \mathbb{R}$ with $c$ being a real number such that the arguments of both Gamma functions have positive real part. Carrying out the above procedure yields the following Mellin--Barnes representation for the box integral
\begin{align}
\phi_4=\frac{N_4}{(2 \pi i)^2} \int\limits_{\kappa_4 +i \mathbb{R}^2} \dd z_1 \wedge \dd z_2 \, \omega_4  ,
\label{eqn:MBboxint}
\end{align}
where
\begin{align}
\omega_4 \! = \! u^{z_1} v^{z_2} \Gamma_{-z_1} \Gamma_{-z_2} \Gamma_{1-\gamma-z_1} \Gamma_{1-\gamma'-z_2} \Gamma_{\alpha+z_1+z_2} \Gamma_{\beta+z_1+z_2} ,
\label{eqn:MBintegrandboxint}
\end{align}
and with $N_4$ as defined in equation \eqref{eqn:BoxNormalizConst}. Here, $\kappa_4$ labels a point in $\mathbb{R}^2$ and is again chosen such that all Gamma function arguments have positive real part. In \Figref{fig:MBPoles} the latter region is depicted as a meshed (blue) triangle in the center of the plot and $\kappa_4$ corresponds to the orange dot, which can be moved within the fundamental triangle without changing the value of the integral. 

The standard method to compute integrals of the form \eqref{eqn:MBbasicformula} and \eqref{eqn:MBboxint} is to use Jordan's lemma in conjunction with the residue theorem \cite{Passare1994}, leading to series representations, which under favorable circumstances can be summed. To apply Jordan's lemma, one first needs to analyze in which domains of the integration space the integrand is a decreasing function. An important quantity in this context is the vector $\vec{\Theta}$ which for denominator-free integrands of the form
\begin{align}
\prod\limits_i \Gamma_{a_{i1} z_1 + \ldots+ b_{in} z_n +c_i} ,
\end{align}
is defined as
\begin{align}
\Theta_j=\sum\limits_i a_{ij} .
\label{eqn:vectheta}
\end{align}
Evaluating this quantity for the integrals \eqref{eqn:MBbasicformula} and \eqref{eqn:MBboxint} shows that both have $\vec{\Theta}=0$ thus corresponding to the so-called degenerate case \cite{zhdanov1998studying}. In this case, the Gamma functions are essentially balanced and integration contours can typically be closed in multiple ways leading to series representations which are valid in different kinematic regimes. For example, the integration contour $C$ in equation \eqref{eqn:MBbasicformula} can be closed via the left or right half-plane resulting in series representations which converge for $|A/B|>1$ and $|A/B|<1$, respectively. Similarly, we will see that multiple series representations of the box integral coexist which are analytic continuations of each other. To obtain these, we proceed by analyzing the singularity structure of the Mellin--Barnes integrand \eqref{eqn:MBintegrandboxint}. The Gamma functions have poles for non-positive integer values of their arguments. In the $\text{Re}(z_1)$-$\text{Re}(z_2)$-plane these poles correspond to singular lines, see \Figref{fig:MBPoles}. For example, $\Gamma_{-z_1}$ has poles for $z_1=m$ with $m$ being a non-negative integer and these are depicted as vertical solid orange lines. Similarly, the Gamma functions involving only $z_2$ lead to the green horizontal lines while those depending on the linear combination of both $z$ variables correspond to the diagonal lines. An interesting point to note is that due to the special structure of the Mellin--Barnes integrand \eqref{eqn:MBintegrandboxint}, finding the zeros of the fundamental solution \eqref{eqn:Rec-sol-1} is essentially the same thing as finding the zeroth representatives of all (infinite) families of singular lines. 

Let us now turn to the question of how to express the Mellin--Barnes integral \eqref{eqn:MBboxint} as a sum of residues. The residues need to be computed at points where singular lines intersect but we have not yet explained which subgroups of poles should be summed to obtain a valid series representation of the original integral. However, for two-dimensional integrals there exists a simple graphical procedure which can be used to find all of these regions \cite{zhdanov1998studying,Friot:2011ic}. The first step consists of drawing an arbitrary cone $R$ with vertex $\kappa_4$ in the $\text{Re}(z_1)$-$\text{Re}(z_2)$-plane. The cone is called compatible with the families of singular lines if each line intersects at most one side of the cone $R$. In \Figref{fig:MBPoles} we have drawn the three cones $R_{\text{I},\text{II},\text{III}}$ (boundaries of the red areas) that are compatible with the six families of singular lines. Once such a compatible cone is found, the integral can be expressed as
\begin{align}
\phi_4= N_4 \sum\limits_{\vec{z}^*\in R_i} \underset{\vec{z}=\vec{z}^*}{\mathrm{res}} \; \omega_4  ,
\end{align}
where the summation ranges over all intersection points that lie inside the compatible cone $R_{i}$. As an example, let us compute the representation that results from summing all residues inside cone $R_{\text{I}}$. Obviously, there are four families of poles in this cone, which can be parametrized as
\begin{align}
&\vec{z}_1^*=(m,n) ,  &&\vec{z}_2^*=(m+1-\gamma,n+1-\gamma')  , \nonumber \\
&\vec{z}_3^*=(m+1-\gamma,n)  , &&\vec{z}_4^*=(m,n+1-\gamma') , 
\end{align}
in complete agreement with table \eqref{Values:xy}. Computing residues at positions $\vec{z}_1^*$ yields
\begin{align}
\underset{\vec{z}=\vec{z}_1^*}{\mathrm{res}} \; \omega_4= \Gamma_{\alpha} \Gamma_{\beta}\Gamma_{1-\gamma}\Gamma_{1-\gamma'} \frac{(\alpha,m+n)(\beta,m+n)}{(\gamma,m)(\gamma',n) \, m! \, n!} u^m v^n ,
\label{eqn:MBfundSol}
\end{align}
where $(\lambda,k)$ is the Pochhammer symbol as defined in \eqref{eqn:F4Def}. The residues at positions $\vec{z}_1^*$ obviously correspond to the \emph{fundamental solution} \eqref{eqn:Rec-sol-1}. Since calculating residues in cone $R_\text{I}$ is essentially trivial, we leave the computation of the other residues to the reader and merely state the final result for the Mellin--Barnes integral \eqref{eqn:MBintegrandboxint}
\begin{align}
&\phi_4=N_4
\Big[\Gamma_\alpha\Gamma_{\beta} \Gamma_{1-\gamma'} \Gamma_{1-\gamma}\, 
g_1(u,v)
 \label{eqn:YDefCrossF4FinalMB} \\
&+\, \Gamma_{1+\alpha-\gamma} \Gamma_{1+\beta-\gamma}\Gamma_{\gamma-1}\Gamma_{1-\gamma'} \, 
g_2(u,v)
\nonumber\\
&+ \, \Gamma_{1+\alpha-\gamma'} \Gamma_{1+\beta-\gamma'} \Gamma_{1-\gamma}\Gamma_{\gamma'-1} \, 
g_3(u,v)
\nonumber \\
&+ \, \Gamma_{2+\beta-\gamma-\gamma'}  \Gamma_{2+\alpha-\gamma-\gamma'}\Gamma_{\gamma'-1}\Gamma_{\gamma-1}\, 
g_4(u,v)\Big] ,\nonumber
\end{align} 
with $g_i(u,v)$ and $N_4$ as defined in \Secref{sec:ParamBox}. Note that the above result is in complete agreement with equation \eqref{eqn:YDefCrossF4Final} and the Mellin--Barnes result for the triangle integral of \cite{Boos:1990rg}. Summing the residues in the other two cones is most conveniently done by performing the change of variables $z_1'=z_1$ and $z_2'=z_1+z_2$ and yields similar expressions but with Appell functions depending on $(u/v,1/v)$ and $(v/u,1/u)$, respectively. The expressions obtained by summing over residues in cone $R_{\text{II}}$ and $R_{\text{III}}$ exactly agree with those obtained by applying the $F_4$-identity \eqref{eq:FuncIdF4} to equation \eqref{eqn:YDefCrossF4FinalMB}, thus showing that all three expressions are indeed analytic continuations of each other which converge in different kinematic regions. The above arguments now also make it clear why we chose to label the cones in exactly the same way as we labeled kinematic regions in  \Secref{sec:recsBox}: the three cones are in one-to-one correspondence with the three kinematic regions in \Figref{fig:ParameterSpace}. 

Finally, let us note that all four families of poles $\{\vec{z}_1^*,\vec{z}_2^*,\vec{z}_3^*,\vec{z}_4^*\}$ in cone $R_{\text{I}}$ individually lead to a Yangian invariant quantity. This statement follows immediately from the discussion in  \Secref{sec:recsBox} and does also apply to the eight remaining families of poles. This shows that in order to obtain a Yangian invariant, one merely needs to sum over all residues originating from the same type of intersection of singular lines, see \Figref{fig:MBPoles}. Summing over all residues inside a given cone is apparently not required for Yangian symmetry.

Having discussed the box integral in great detail, let us now turn to the $D$-dimensional hexagon integral \eqref{eq:HexagonDefConformal}. Applying nine times the Mellin--Barnes identity to the Feynman parameter representation \eqref{eqn:PhisFeynPara} yields
\begin{align}
\phi_6=\frac{N_6}{(2 \pi i)^9} \int\limits_{\kappa_6 +i \mathbb{R}^9} \dd z_1 \wedge \ldots \wedge \dd z_9 \, \omega_6  ,
\label{eqn:MBHexint}
\end{align}
where
\begin{align}
N_6=\frac{\pi^{2+\alpha+\beta_1+\beta_2+\beta_3-\gamma_1-\gamma_2}}{\Gamma_{\beta_1} \Gamma_{\beta_2} \Gamma_{\beta_3} \Gamma_{1+\alpha-\gamma_1} \Gamma_{1+\alpha -\gamma _2} \Gamma_{2+\beta_1+\beta_2+\beta_3-\gamma_1-\gamma_2}} ,
\end{align}
and
\begin{align}
\omega_6 =&  
	\prod\limits_{i=1}^9 \brk*{ w_i^{z_i} \Gamma_{-z_i} }  \, 
	\Gamma_{\alpha+z_1+z_2+z_3+z_4+z_5+z_6+z_7+z_8+z_9}  \nonumber \\
&\times \Gamma_{\beta_1+z_1+z_5+z_8+z_9} \Gamma_{\beta_2+z_2+z_6+z_7+z_8} \nonumber \\
&\times \Gamma_{\beta_3+z_3+z_4+z_5+z_6} \Gamma_{1-\gamma_2-z_4-z_5-z_6-z_7-z_8-z_9} \nonumber \\
&\times \Gamma_{1-\gamma_1-z_1-z_2-z_3-z_5-z_6-z_8} ,
\end{align}
with the cross ratios $w_i$ as defined in \Appref{sec:CrossRatios}. In complete analogy with the two-dimensional case, the vector $\kappa_6 \in \mathbb{R}^9$ is defined such that the arguments of all Gamma functions have positive real part. 

Evaluating the vector $\vec{\Theta}$ as defined in equation \eqref{eqn:vectheta} shows that the above integral is degenerate as well, i.e. $\vec{\Theta}=0$, so that presumably multiple series representations coexist. Since the integration space is nine-dimensional, the graphical method outlined above can no longer be applied and one needs to rely on purely algebraic methods to find all compatible cones. This, however, is left for future work. Instead, we will content ourselves with computing the analog of the fundamental solution \eqref{eqn:MBfundSol}. For this, we only need to find the residues at the intersection points $\vec{z}_1^*=(n_1,n_2,n_3,n_4,n_5,n_6,n_7,n_8,n_9)$. We obtain
\begin{align}
& \underset{\vec{z}=\vec{z}_1^*}{\mathrm{res}} \; \omega_6 = \Gamma_{\alpha} \Gamma_{\beta_1} \Gamma_{\beta_2} \Gamma_{\beta_3} \Gamma_{1-\gamma_1} \Gamma_{1-\gamma_2} (-1)^{n_5+n_6+n_8} \nonumber \\
& \hspace{5mm}
\times\frac{(\alpha,M_1)(\beta_1,M_2)(\beta_2,M_3)(\beta_3,M_4)}{(\gamma_1,M_5)(\gamma_2,M_6)} \prod\limits_{i=1}^9 \frac{w_i^{n_i}}{n_i!}  ,
\end{align}
where the $M_{j=1,\dots,6}$ were defined in \eqref{eq:MHexa} and again we used Pochhammer notation to emphasize the hypergeometric nature of the residues. More precisely, summing over all the residues $\vec{z}_1^*$ yields a multivariate hypergeometric series of Srivastava--Daoust type, see \cite{srivastava1969certain,srivastava1985multiple}. Picking other sets of residues leads to similar expressions with some linear combinations of $n_i$'s replaced by others, all of them representing individual Yangian invariants.

Let us finish this section with a remark on the consideration of $D$-dimensional integrals with 
deformed propagator powers. While the deformation naively just adds another layer of complexity, it actually turns out to be a blessing in the context of Mellin--Barnes integrals as it disentangles different sets of poles which would otherwise overlap. For the box integral the latter statement becomes transparent by comparing the result \eqref{eqn:YDefCrossF4FinalMB} to the result for the undeformed box integral \eqref{eqn:BoxZeroDefF4}. In case of the Hexagon integral it even seems that the deformation is what makes the residue theorem applicable in the first place since in the undeformed case there exist singular points in which more than nine singular hyperplanes intersect, thus making the residues a priori no longer well-defined.

\section{Conclusions and Outlook}

In this paper we have demonstrated that it is possible to bootstrap Feynman integrals using their Yangian symmetry.
In the case of the 2-variable box integral we have shown in full detail that the Yangian constrains the functional form of the integral to a space spanned by four Appell hypergeometric functions $F_4$. Their linear combination is fixed through the integral's permutation symmetries, and the overall constant is determined by relating the integral to the star-triangle relation in a coincidence limit of external points. Hence, we have completely bootstrapped the $D$-dimensional conformal box integral with generic propagator powers. 
For the much harder 9-variable hexagon and double box integrals, we have discussed the analogous Yangian constraints, which in each case can be translated into a system of 15 differential equations in the conformal cross ratios. For the hexagon PDEs we have argued that these constraints are solved by a set of 2530 generalized Lauricella series. Due to this large number and the poor understanding of the convergence properties of these solutions, it was not possible to identify a linear combination of these series that corresponds to the integral. 

These investigations suggest plenty of directions that require further understanding.
Firstly, the discussion in \Secref{sec:MB} illustrates the close connection to the Mellin--Barnes technique for the computation of Feynman integrals, which in turn can be understood through the close connection between Mellin--Barnes integrals and hypergeometric functions. In our context, the Mellin--Barnes integrand is closely related to the fundamental solution of the Yangian recurrence equations.
In fact, both approaches share similar problems for integrals with a larger number of variables. These are to identify the correct linear combination of series solutions and to understand their convergence properties. Already in the 2-variable case a proper convergence analysis is laborious, see e.g.\ \cite{Friot:2011ic} for the explicit discussion of the convergence of 2-variable Mellin--Barnes integrals. This underlines the importance of getting better control over the mathematical properties of the often poorly understood multi-variable generalizations of hypergeometric functions. A serious convergence analysis for the 9-variable case seems indeed very hard. 
\bigskip

Let us point out that in this paper we have observed that the Yangian differential equations for Feynman integrals 
can be formulated for generic spacetime dimension $D$, whereas the symmetry found in \cite{Chicherin:2017frs} was phrased in different but fixed spacetime dimensions $D=3,4,6$. 
While here the approach with generic $D$ emerged naturally, the case most interesting for phenomenological applications is $D=4$ with unit propagator powers. It is thus natural to ask whether working directly in this limit, the considered integrals can be bootstrapped more easily. For the case of the box integral we have seen in \Secref{sec:BWfromY} that indeed in this limit the Yangian constraints yield the solution by elementary means. This solution, however, seems less algorithmic than the bootstrap for generic propagator powers and thus less simple to generalize. The subtleties of the limit of unit propagator powers discussed in \Secref{sec:ParamBox} show that it has various advantages to work with the deformed integral. Nevertheless it is clearly interesting to further investigate the unit-propagator bootstrap, e.g.\ by studying the Yangian constraints on the symbol of the respective function, similar to the approach of \cite{DelDuca:2011wh} for the hexagon integral with three massless and three massive corners.

In addition to the Yangian constraints, here we used the permutation symmetries of the box integral as well as a coincidence limit of two external legs to fix it. Aesthetically it would be more pleasing to fix an integral by integrability (alias Yangian symmetry) alone. That this is indeed in reach is suggested by the recursive structure described in \Secref{sec:RecStruc}. Similar to the conformal symmetry of scattering amplitudes in $\mathcal{N}=4$ super Yang--Mills theory \cite{Bargheer:2009qu}, it may be possible to include this structure into the representation of the Yangian on Feynman integrals. Certainly in the case of on-shell legs, the conformal differential equations acquire inhomogeneities, and the resulting equations have been shown to yield powerful tools for the computation of Feynman integrals \cite{Chicherin:2017bxc,Chicherin:2018ubl}.

For the double box, the case with unit propagator powers and on-shell legs is known to be described by elliptic functions, whose theory in the context of Feynman integrals is still under construction
\cite{CaronHuot:2012ab,Bourjaily:2017bsb,Adams:2018bsn,Broedel:2017kkb}. 
It would be very interesting to apply the Yangian PDEs studied in this paper to an ansatz for this integral, once such an ansatz becomes available. Moreover, explicitly relating the above elliptic formalism and the hypergeometric building blocks that naturally emerge in the context of the Yangian PDEs should be a worthwhile goal.

The relation of Yangian symmetry to the PDEs \eqref{eq:PDEeq0} shows that the roots of the Yangian symmetry lie in systems of partial differential equations in the conformal cross ratios.
Notably, there are various strategies to write down differential equations for Feynman integrals. In particular, the more formal approach of the recent papers \cite{delaCruz:2019skx,Klausen:2019hrg,Feng:2019bdx} using the Gelfand--Kapranov--Zelevinsky systems seems closely related to ours. It would be interesting to study the systematics behind this relation and to see in how far the resulting systems of differential equations agree.

The main tool of the present paper is the Yangian Hopf algebra acting on certain Feynman graphs. Recently, similar algebraic structures were found in the context of other classes of Feynman integrals, in particular also in the context of Appell and Lauricella hypergeometric functions, see e.g.\ \cite{Brown:2019jng,Abreu:2019wzk}. It should be enlightening to investigate the parallels in these approaches and to understand whether both algebraic structures coincide or coexist.

Curiously, integrability also enters the scene of conformal correlation functions from a different direction. In \cite{Isachenkov:2016gim} and several follow-up works it has been shown that conformal blocks can be understood as eigenfunctions of an integrable Calogero--Sutherland Hamiltonian. There the eigenvalue equation is obtained from the conformal Casimir equation known to hold for the conformal blocks. Understanding the connection between that approach and the integrability properties of conformal correlators employed in the present paper should be instructive. A natural starting point is the box integral considered here, which can be interpreted as a correlation function in the fishnet theory of \cite{Gurdogan:2015csr}.

While the present paper deals with the constraints for scalar Feynman integrals, also Feynman integrals including fermions can be shown to obey a Yangian symmetry \cite{Chicherin:2017frs}. The respective diagrams are again interpreted as correlators in a generalized fishnet model, see also \cite{Caetano:2016ydc,Kazakov:2018gcy}. An obvious task is thus to bootstrap the simplest examples of fermionic Feynman integrals and to see how far this approach reaches for those cases.

Certainly, it is an interesting question on its own to understand the constraining power of integrability in the context of four-dimensional high energy physics. However, a more ambitious goal of this program is to develop efficient integrability methods for the computation of Feynman integrals and to understand the deeper mathematical structures underlying quantum field theory. Here it will be important to further extend the traditional integrability toolbox to the situations at hand. For the case of the yet unknown six-point integrals discussed in this paper, the present status report furnishes the groundwork for further progress into this direction.


\acknowledgments{
We thank 
Lance Dixon, J\"urg Kramer, Julian Miczajka and Cristian Vergu for helpful discussions. Moreover, we are grateful to Julian Miczajka for comments and suggestions on the manuscript. DM and FL thank the Pauli Center and ETH Z{\"u}rich for support and hospitality. The work of HM has been supported by the grant no.\ 615203 from the European Research Council under the 
FP7 and by the Swiss National Science Foundation through the NCCR SwissMAP. DM was supported by DFF-FNU through grant number DFF-FNU 4002-00037. The work of FL is funded by the Deutsche Forschungsgemeinschaft (DFG, German Research Foundation)--Projektnummer 363895012. 
}
\appendix

\section{Details on Conformal Cross Ratios}
\label{sec:CrossRatios}
In order to understand the degrees of freedom that remain after imposing 
conformal invariance, it is helpful to consider the conformal 
compactification of our underlying spacetime, on which the conformal 
group acts linearly \cite{Dirac:1936fq}. 
In the case of Euclidean four-dimensional space, 
the conformal compactification can be realized by the light cone 
in $\mathbb{R}^{(1,5)}$, 
\begin{align}
- \big( X^{0'}\big) ^2 
+ \sum \limits _{i=1} ^5 \brk*{X^i}^2 = 0 , 
\label{eq:DiracCone}
\end{align}
modulo the identification $X \sim \lambda X$. We can map this space 
to $\mathbb{R}^4$ and vice versa employing the identifications
\begin{align}
x^\mu &= \frac{X^\mu}{X^0 + X^5} , & 
\brk[s]*{X} =  \brk[s]*{1+x^2: 2 x^\mu : 1-x^2} . 
\label{map:DiracCone}
\end{align} 
We consider six points $\lbrace X_i \rbrace$ and constrain their 
coordinates as far as possible using $\mathrm{SO}(1,5)$ transformations, 
i.e.\ conformal symmetry. 
For the first two points, it is clear that using only $\mathrm{SO}(5)$ 
rotations, we can reach the form 
\begin{align}
\brk[s]*{X_1} &= \brk[s]*{ 1: 0 : 0 : 0 : 0 : 1 } , \\ 
\brk[s]*{X_4} &= \brk[s]*{ a: b : 0 : 0 : 0 : c } . \nonumber
\end{align}
We can then determine the stabilizer of $\brk[s]*{X_1} $ by requiring 
that $M \in \mathfrak{so}(1,2)$ only acts as a scaling on 
$y=(1,0,1)$ and exponentiation of the elements spanning this space. 
In this way, we find that we can reach the form 
\begin{align}
\brk[s]*{X_4} &= \brk[s]*{ 1: 0 : 0 : 0 : 0 : -1 } , 
\label{eq:Configs}
\end{align}
corresponding to infinity in $\mathbb{R}^4$. Proceeding similarly, we 
find that $ \brk[s]*{X_3}$ can be brought to the form 
\begin{align}
\brk[s]*{X_3} = \brk[s]*{ 1: 1 : 0 : 0 : 0 : 0 } , 
\end{align}
which leaves us with $\mathrm{SO}(3)$ as the stabilizer of these three 
points. It is then straight-forward to constrain the following three 
points to 
\begin{align}
x^\mu _2 &= 
	(z_1 , z_2 , 0 , 0 ) , \\
x^\mu _5 &= 
	(z_3 , z_4 , z_5 , 0 ) , \\
x^\mu _6 &= 
	(z_6 , z_7 , z_8 , z_9 ) ,		
\end{align}
the points on the Dirac light cone \eqref{eq:DiracCone}
follow from the relation \eqref{map:DiracCone}. 
It becomes clear from the above construction that in the case of 
four points, there are two degrees of freedom (compared to the 
$16-15=1$ one could expect based on the dimension of the conformal 
group), since a stabilizer group $\mathrm{SO}(2)$ remains. 
We also note that the range of the 
coordinates $z_i$ is clear, since we can always pick the respective 
points in $\mathbb{R}^4$ they represent. However, by performing rotations 
with angle $\pi$ in $\mathbb{R}^4$, we can always enforce that 
\begin{align}
z_2 & \geq 0 , & 
z_5 & \geq 0 , & 
z_9 & \geq 0 . 
\end{align}

In terms of the vectors $\brk[s]*{X_i}$ on the Dirac cone, the conformal 
cross ratios are given by 
\begin{align}
u_{ijkl} = \frac{(X_i \cdot X_j) (X_k \cdot X_l)}
	{(X_i \cdot X_k) (X_j \cdot X_l) } 
	= \frac{x_{ij}^2 x_{kl}^2 }
	{x_{ik}^2  x_{jl}^2  } . 
\end{align}
It is helpful, to express the $z$-variables in terms of one set of 
independent cross ratios. This facilitates to check whether any other set 
of cross ratios is independent (by expressing it in terms of the first set) 
and is also a handy tool in order to derive the relations between two given 
sets of cross ratios. 
For this purpose, we consider the following set of cross ratios  
\begin{align}
v_1 &= u_{1234} = z_1^2 + z_2 ^2 , \nonumber \\
v_2 &= u_{1432} = (z_1 - 1)^2 + z_2 ^2 ,\nonumber \\ 
v_3 &= u_{1435} = (z_3 - 1)^2 + z_4 ^2  + z_5 ^2 ,\nonumber \\
v_4 &= u_{1534} = z_3^2 + z_4 ^2 + z_5 ^2 , \nonumber \\ 
v_5 &= u_{1234} u_{1425} 
	= (z_1 - z_3)^2 + (z_2 - z_4) ^2  + z_5 ^2 , 
	\label{eq:9OriginalCrossRatios}\\
v_6 &= u_{1436} = (z_6 - 1)^2 + z_7 ^2 + z_8 ^2 + z_9 ^2, \nonumber\\
v_7 &=u_{1634} = z_6^2 + z_7 ^2 + z_8 ^2 + z_9 ^2 , \nonumber\\
v_8 &=u_{1234} \, u_{1426} \nonumber \\
	&= (z_1 - z_6)^2 + (z_2 - z_7) ^2  + z_8 ^2 + z_9 ^2 , \nonumber \\
v_9 &=u_{1534} \, u_{1456} \nonumber \\
	&= (z_3 - z_6)^2 + (z_4 - z_7) ^2  + (z_5 - z_8) ^2 + z_9 ^2 \nonumber . 	
\end{align}
For the first two of the above cross ratios, we also employ the notation
\begin{align}
v_1 &= z \bar z \equiv u , &
v_2 &= (1-z) (1- \bar z) \equiv v , 
\end{align}
with $z = z_1 + i z_2$ and $\bar z$ its complex conjugate. For these, we 
note the relations 
\begin{align}
z_1 &= \half (1+u-v) , &
z_2 &= \sqrt{u - z_1 ^2} , 
\label{eq:uvdom}
\end{align}
from which we read off that $u,v$ are restricted to the domain 
\begin{align}
4u \geq (1+u-v)^2 , 
\end{align}
in which the radicand in \eqref{eq:uvdom} is non-negative. Working with a 
Minkowskian signature, we could reach the configuration 
\begin{align}
\brk[s]*{X_1} &= \brk[s]*{ 1: 0 : 0 : 0 : 0 : 1 } , \nonumber \\
\brk[s]*{X_2} &= \brk[s]*{ 1-\tilde{z}_1 \tilde{z}_2: \tilde{z}_1+\tilde{z}_2 : 
	\tilde{z}_1-\tilde{z}_2 : 0 : 0 : 1 +\tilde{z}_1 \tilde{z}_2 } , \nonumber \\
\brk[s]*{X_3} &= \brk[s]*{ 0: 1 : 0 : 0 : 0 : 1 } , \\
\brk[s]*{X_4} &= \brk[s]*{ 1: 0 : 0 : 0 : 0 : -1 } , \nonumber
\end{align}
which leads to the expressions 
\begin{align}
\tilde{u} &= \tilde{z}_1 \tilde{z}_2, & 
\tilde{v} &= (1-\tilde{z}_1)(1-\tilde{z}_2) .  
\end{align}
Solving these for $\tilde{z}_i$ shows that these cross ratios are 
restricted by the relation
\begin{align}
4 \tilde{u} \leq (1+\tilde{u}-\tilde{v})^2 , 
\end{align}
covering the opposite domain of the Euclidean cross ratios and overlapping 
only along the line $4u= (1+u-v)^2$. 

Returning to the Euclidean cross ratios given in \eqref{eq:9OriginalCrossRatios}, 
we note that expressing the $z$-variables in terms of these cross ratios is 
a straight-forward exercise (which incidentally also shows that the given 
cross ratios are indeed independent). We find the relations (in addition 
to the ones given above for $z_1$ and $z_2$)
\begin{align}
z_3 &= \half (1 + v_4 - v_3) , \nonumber\\
z_4 &= \tfrac{1}{2z_2} \brk*{v_1 - 2 z_1 z_3 + v_4 - v_5 } , \nonumber\\
z_5 &= \sqrt{v_4 - z_3 ^2 - z_4 ^2} , \nonumber\\
z_6 &= \half ( 1 + v_7 - v_6 ) , \\
z_7 &= \tfrac{1}{2z_2} \brk*{v_1 - 2 z_1 z_6 + v_7 - v_8} , \nonumber\\
z_8 &= \tfrac{1}{2z_5} \brk*{v_4 - 2 z_3 z_6 - 2 z_4 z_7 + v_7 - v_9 } , \nonumber\\
z_9 &= \sqrt{v_7 - z_6^2 - z_7 ^2 - z_8^2} . \nonumber
\end{align}
The cross ratios $v_i$ given in this appendix merely serve as a tool to 
understand the range of cross ratios, their independence and to establish 
relations between competing sets of cross ratios. In the discussion 
of the double box and hexagon, respectively, we employ different sets 
of cross ratios. We list the explicit definition of these below:
\begin{align}
u_1&=\frac{x_{14}^2x_{23}^2}{x_{13}^2x_{24}^2},
&
u_2&=\frac{x_{15}^2x_{24}^2}{x_{14}^2x_{25}^2},
&
u_3&=\frac{x_{16}^2x_{25}^2}{x_{15}^2x_{26}^2},
\nonumber\\
u_4&=\frac{x_{25}^2x_{34}^2}{x_{24}^2x_{35}^2},
&
u_5&=\frac{x_{26}^2x_{35}^2}{x_{25}^2x_{36}^2},
&
u_6&=\frac{x_{12}^2x_{36}^2}{x_{13}^2x_{26}^2},
\nonumber\\
u_7&=\frac{x_{36}^2x_{45}^2}{x_{35}^2x_{46}^2},
&
u_8&=\frac{x_{13}^2x_{46}^2}{x_{14}^2x_{36}^2},
&
u_9&=\frac{x_{14}^2x_{56}^2}{x_{15}^2x_{46}^2}.
\label{eq:9crosscr}
\end{align}
For completeness we also re-display the redefined cross ratios employed in the context of the hexagon:
\begin{align}
w_1&=u_3,
&
w_2&=u_3 u_5,
&
w_3&=u_3u_5u_7,
\nonumber\\
w_4&=u_9,
&
w_5&=u_2u_3u_9,
&
w_6&=u_2u_3u_4u_5u_9,
\nonumber\\
w_7&=u_8u_9,
&
w_8&=u_1u_2u_3u_8u_9,
&
w_9&=u_6u_8u_9.
\end{align}

The introduction of the $z$-variables is also helpful in order to 
discuss the `independence' of the vectors 
\begin{align}
a_{jk} ^\mu &= \frac{x_{jk}^\mu}{x_{jk}^2} , 
\end{align}
which allows us to conclude that the equation
\begin{align}
\sum \limits _{j<k} a_{jk} ^\mu \, \mathrm{PDE}_{jk} \phi =0
\label{eq:InvCond}
\end{align}
implies that 
\begin{align}
\mathrm{PDE}_{jk} \phi =0 \quad \text{for all } j,k. 
\label{eq:PDEInv}
\end{align}
In order to see this, we start from the explicit configurations 
$X_i$ given around equation \eqref{eq:Configs} and employ a special 
conformal transformation, which is represented by
\begin{align}
\Lambda_c = \begin{pmatrix}
1 + \half c^2 & - c^\mu & -\half c^2 \\
-c^\mu & \mathbf{1}_4 & c^\mu \\
\half c^2 & - c^\mu & 1 - \half c^2 
\end{pmatrix}
\end{align}
on the conformal compactification, cf.\ e.g.\ \cite{Dorn:2012cn}, also for a more 
detailed discussion on the conformal compactification. 
Additionally, it is helpful to employ a rotation with angle $\pi$ in the 
(5,6)-plane in order to avoid one of the points being mapped to infinity 
in Euclidean $\mathbb{R}^4$. 
In this way, we obtain explicit expressions for the $a_{ij}^\mu$, which 
depend on variables $z$ and the parameters $c^\mu$ of the special conformal 
transformation. We note that equation \eqref{eq:InvCond} is constructed in
such a way that the coefficients $\mathrm{PDE}_{jk} \phi$ are conformally 
invariant and thus independent of the parameters $c^\mu$. 
Expanding all four components of \eqref{eq:InvCond} in powers of the 
parameters $c^\mu$ is then sufficient to establish \eqref{eq:PDEInv}. 

\section{Shift Identities for the Box}
\label{sec:shifts}

The expansion coefficients \eqref{eqn:Rec-sol-1} of the solutions to 
the Appell PDEs,
\begin{align*}
g_{mn} ^{\alpha \beta \gamma \gamma'} = \frac{1}{\Gamma_{m+1}\Gamma_{n+1} \Gamma_{m+\gamma}
	\Gamma_{n+\gamma'} \Gamma_{1-m-n-\alpha}\Gamma_{1-m-n-\beta}} , 
\end{align*}
satisfy a number of shift identities such as \eqref{eq:shift1} and \eqref{eq:OneShiftg} given above.
For a more systematic understanding, note that all 48 possible shifts of this type can be generated 
from the following three relations: 
\begin{align}
g_{mn} ^{\alpha \beta \gamma \gamma'} &= 
	g_{n, m+1-\gamma} ^{1+\beta-\gamma, 1+\alpha-\gamma, \gamma', 2 - \gamma} , \nonumber \\
g_{mn} ^{\alpha \beta \gamma \gamma'} &= 
	g_{m, -m -n -\beta} ^{1+\beta-\gamma', \beta, \gamma, 1+\beta-\alpha} , \\
g_{mn} ^{\alpha \beta \gamma \gamma'} &= g_{mn} ^{\beta \alpha \gamma \gamma'} . \nonumber
\end{align}
The first two of these shifts correspond to the generators $\sigma_1, \sigma_2$ of the permutation 
group, respectively. 
In order to derive relations such as \eqref{eq:SumRelation1} for all 12 series representations of the 
solutions of the Appell PDEs, we note the following shifts:  
\begin{align}
g_{m+1-\gamma,n} ^{\alpha \beta \gamma \gamma'} &= 
	g_{mn} ^{1+\alpha-\gamma,1+\beta-\gamma,2-\gamma,\gamma'} , \nonumber \\
g_{m,n+1-\gamma'} ^{\alpha \beta \gamma \gamma'} &= 
	g_{mn} ^{1+\alpha-\gamma',1+\beta-\gamma',\gamma,2-\gamma'} , \nonumber \\	
g_{m+1-\gamma,n+1-\gamma'} ^{\alpha \beta \gamma \gamma'} &= 
	g_{mn} ^{2+\alpha-\gamma-\gamma',2+\beta-\gamma-\gamma',2-\gamma,2-\gamma'} , \nonumber \\	
g_{m-\alpha,n} ^{\alpha \beta \gamma \gamma'} &= 
	g_{-m-n,n} ^{\alpha,1+\alpha-\gamma,1+\alpha-\beta,\gamma'} ,	\nonumber \\	
g_{m-\beta,n} ^{\alpha \beta \gamma \gamma'} &= 
	g_{-m-n,n} ^{1+\beta-\gamma,\beta,1+\beta-\alpha,\gamma'} ,	\\	
g_{m+\gamma'-\alpha-1,n+1-\gamma'} ^{\alpha \beta \gamma \gamma'} &= 
	g_{-m-n,n} ^{1+\alpha-\gamma',2+ \alpha -\gamma -\gamma',1+\alpha -\beta,2-\gamma'} , \nonumber \\
g_{m+\gamma'-\beta-1,n+1-\gamma'} ^{\alpha \beta \gamma \gamma'} &= 
	g_{-m-n,n} ^{1+\beta-\gamma',2+ \beta -\gamma -\gamma',1+\beta-\alpha,2-\gamma'}, \nonumber \\	
g_{m,n-\alpha} ^{\alpha \beta \gamma \gamma'} &= 
	g_{m,-m-n} ^{\alpha,1+\alpha-\gamma',\gamma, 1+\alpha-\beta} ,	\nonumber \\	
g_{m,n-\beta} ^{\alpha \beta \gamma \gamma'} &= 
	g_{m,-m-n} ^{1+\beta-\gamma',\beta,\gamma,1+\beta-\alpha} ,	\nonumber \\	
g_{m+1-\gamma,n+\gamma-\alpha-1} ^{\alpha \beta \gamma \gamma'} &= 
	g_{m,-m-n} ^{1+\alpha-\gamma,2+ \alpha -\gamma -\gamma',2-\gamma,1+\alpha -\beta} , \nonumber \\
g_{m+1-\gamma,n+\gamma-\beta-1} ^{\alpha \beta \gamma \gamma'} &= 
	g_{m,-m-n} ^{1+\beta-\gamma,2+ \beta -\gamma -\gamma',2-\gamma,1+\beta-\alpha} . \nonumber			
\end{align}

\section{Feynman Parametrizations}
\label{sec:feynpara}
In this appendix we list (dual conformal) Feynman parameter representations \cite{Hodges:2010kq,SimmonsDuffin:2012uy,Bourjaily:2019jrk} for the integrals discussed in this paper. In their most general form, the integrals read
\begin{align}
I_{4}&= \int  \frac{\dd^D x_0 }{x_{10}^{2a}x_{20}^{2b}x_{30}^{2c} x_{40}^{2d}} = V_4 \,\phi_4\, , \\
I_{3,3}&=\int  \frac{\dd^D x_0 \dd^D x_{0'} }{x_{10}^{2a} x_{20}^{2b} x_{30}^{2c} x_{00'}^{2 \ell} x_{40'}^{2d} x_{50'}^{2e} x_{60'}^{2f}} = V_{3,3} \, \phi_{3,3} \, ,\nonumber \\
I_{6}&= \int  \frac{\dd^D x_0 }{x_{10}^{2a}x_{20}^{2b}x_{30}^{2c} x_{40}^{2d}x_{50}^{2e}x_{60}^{2f}}=  V_{6} \, \phi_{6} \, , \nonumber
\end{align}
where $x^{2a}$ is short for $(x^2)^a$. Note that we evaluate all integrals at their conformal point, i.e.\ the propagator weights at each vertex have to add up to the dimension $D$. The prefactors $V_i$ carry the conformal weight of the integral and are given by
\begin{align}
V_4 &= x_{13}^{2d-D} x_{14}^{2b+2c-D} x_{24}^{-2b} x_{34}^{D-2c-2d} \, , \\
V_{3,3} &= x_{13}^{2 \ell-D} x_{14}^{D-2\ell} x_{15}^{-2d-2e} x_{16}^{2d+2e-2a} x_{26}^{-2b} x_{36}^{D-2c-2\ell} \nonumber\\
&\hspace{14pt}  x_{46}^{2\ell-2d-D} x_{56}^{2d} \, , \nonumber \\
V_6&= x_{15}^{2f-D} x_{16}^{D-2a-2f}x_{26}^{-2b} x_{36}^{-2c}x_{46}^{-2d}x_{56}^{D-2e-2f} \, . \nonumber
\end{align}
The above way of factorizing the integrals leads to the following conformally invariant functions of $2$ and $9$ cross ratios, respectively:
\begin{align}
\label{eqn:PhisFeynPara}
\phi_4&=Q_4 \int\limits_0^\infty \dd\beta_2 \dd\beta_3 \frac{\beta_2^{b-1} \beta_3^{c-1}}{ X_1^{D/2-d} Z_2^d} \, ,\\
\phi_{3,3}&=Q_{3,3} \int\limits_{0}^{\infty} \dd\beta_2 \dd\beta_3 \dd\beta_4 \dd\beta_5 \frac{\beta_2^{b-1} \beta_3^{c-1} \beta_4^{d-1} \beta_5^{e-1}}{X_2^{D/2-\ell} Y^{D/2-f} Z_4^{f}} \, , \nonumber\\
\phi_6&= Q_6 \int\limits_{0}^{\infty} \dd\beta_2 \dd\beta_3 \dd\beta_4 \dd\beta_5 \frac{\beta_2^{b-1} \beta_3^{c-1} \beta_4^{d-1} \beta_5^{e-1}}{Y^{D/2-f} Z_4^{f}} \, , \nonumber
\end{align}
where
\begin{align}
Q_4&=\frac{\pi^{D/2} \Gamma_{D/2-d}}{\Gamma_a \Gamma_b \Gamma_c} \, , \\
Q_{3,3}&=\frac{\pi^D \Gamma_{D/2-\ell} \Gamma_{D/2-f}}{\Gamma_a \Gamma_b  \Gamma_c \Gamma_{\ell}\Gamma_d \Gamma_e}  \, ,\nonumber \\
Q_6&=\frac{\pi^{D/2} \Gamma_{D/2-f}}{\Gamma_a \Gamma_b  \Gamma_c \Gamma_d \Gamma_e} \, \nonumber ,
\end{align}
and
\begin{align}
X_1&=\beta_2 u + \beta_3  +\beta_2 \beta_3 v \, , \\
X_2&=\beta _2 u_6 u_9+\beta _3 u_9+\beta _2 \beta _3 u_1 u_2 u_3 u_9\, , \nonumber\\
Y&=u_8 X_2 +\beta _4 u_9 +\beta _2\beta _4  u_2 u_3u_9+\beta _3 \beta _4 u_2 u_3 u_4 u_5u_9  \nonumber\\
&\quad +\beta _5+\beta _2 \beta _5 u_3+\beta _3 \beta _5 u_3 u_5+\beta _4 \beta _5 u_3 u_5 u_7 \, ,\nonumber\\
Z_i&=1+\beta _2+\beta _3+ \ldots +\beta _{i+1} \, ,\nonumber
\end{align}
with the cross ratios as defined in \Appref{sec:CrossRatios}. 
Note that the above way of parametrizing the integrals makes the differential equation \eqref{eqn:PSVdiffeq} fairly obvious. Indeed, by noting that $\partial_{u_8} Y= X_2$, one readily concludes that for $D/2-\ell=1$ the following relations holds
\begin{align}
\partial_{u_8} \phi_{3,3}(D) =-\frac{\pi^{D/2-1}}{\Gamma_{\ell}}\phi_6(D+2) \, .
\end{align} 
\section{Functional Identities for the Double Box}
\label{sec:FuncIdentitiesDoubleBox}
In addition to the invariance under the permutations \eqref{eq:DBPerms}, the conformal function $\phi_{3,3}$ for the double box integral fulfills the following functional identities corresponding to the given transpositions of the six external legs of the integral:
\begin{allowdisplaybreaks}
\begin{widetext}
\begin{align}
&(12):
&
&\phi_{3,3}\brk!{\sfrac{1}{u_1}, \sfrac{1}{u_2},\sfrac{1}{u_3},u_2 u_4,u_3 u_5,\sfrac{u_6}{u_1 u_2 u_3},u_7,u_1 u_8,u_2 u_9}
=u_1 u_2 \,\phi_{3,3}\brk!{u_1,u_2,u_3,u_4,u_5,u_6,u_7,u_8,u_9},
\\
&(13):
&
&\phi_{3,3}\brk!{u_4 u_5 u_6, \frac{1}{u_4},\sfrac{1}{u_5},\sfrac{1}{u_2},\sfrac{1}{u_3},u_1 u_2 u_3,u_3 u_5 u_7,\sfrac{u_8}{u_2 u_3 u_4 u_5},u_2 u_4 u_9}
=\phi_{3,3}\brk!{u_1,u_2,u_3,u_4,u_5,u_6,u_7,u_8,u_9} ,
\\
&(23):
&
&\phi_{3,3}\brk!{\sfrac{u_1}{u_4 u_5 u_6}, u_2 u_4, u_3 u_5,\sfrac{1}{u_4},\sfrac{1}{u_5}, \sfrac{1}{u_6} , u_5 u_7, u_6 u_8, u_9}
=u_5 u_6 \,\phi_{3,3}\brk!{u_1,u_2,u_3,u_4,u_5,u_6,u_7,u_8,u_9},
\\
&(46):
&
&\phi_{3,3}\brk!{u_1 u_2 u_3, \sfrac{1}{u_3}, \sfrac{1}{u_2},\sfrac{1}{u_5},\sfrac{1}{u_4}, u_4 u_5 u_6 , u_2 u_4 u_9, \sfrac{u_8}{u_2 u_3 u_4 u_5}, u_3 u_5 u_7}
= \phi_{3,3}\brk!{u_1,u_2,u_3,u_4,u_5,u_6,u_7,u_8,u_9},
\\
&(45):
&
&\phi_{3,3}\brk!{u_1 u_2 , \sfrac{1}{u_2}, u_2 u_3 ,\sfrac{1}{u_4}, u_4 u_5, u_6 , \sfrac{u_7}{u_2 u_4 u_9} , u_8 u_9, \sfrac{1}{u_9}}
= u_2 u_9 \,\phi_{3,3}\brk!{u_1,u_2,u_3,u_4,u_5,u_6,u_7,u_8,u_9},
\\
&(56):
&
&\phi_{3,3}\brk!{u_1 ,u_2 u_3 , \sfrac{1}{u_3}, u_4 u_5 ,\sfrac{1}{u_5}, u_5 u_6, \sfrac{1}{u_7} , u_7 u_8 , \sfrac{u_9}{u_3 u_5 u_7}}
= u_5 u_7 \,\phi_{3,3}\brk!{u_1,u_2,u_3,u_4,u_5,u_6,u_7,u_8,u_9} .
\end{align}
\end{widetext}
\end{allowdisplaybreaks}

\section{Hexagon PDEs}
\label{sec:hexPDEs}

We list the PDEs for the conformal hexagon function that follow from Yangian symmetry. We use the notation 
$\theta_{i_1 \ldots i_n} = \theta_{i_1} + \ldots +\theta_{i_n}$ and we 
abbreviate the sum of all Euler operators by
$\theta_\Sigma = \theta_1 + \ldots + \theta_9$: 
\begin{allowdisplaybreaks}
\begin{widetext} 
\begin{align}
\mathrm{PDE}_{12} &= 
	w_1 w_5 w_9 \brk[s]*{ w_4 \theta_7 + w_7 \theta_4 + w_4 w_7 \left(\theta _\Sigma + \alpha \right) }
		\left(\theta _{1 5 8 9} + \beta _1 \right) 
	-w_1 w_4 w_6 w_9 \theta _5 \theta _7 \nonumber \\
& \quad -w_3 w_5 w_7 w_9 \theta _1 \theta _4 
	-w_2 w_4 w_5 w_9 \theta _1 \theta _7
	+w_1 w_4 w_5 w_7 \theta _9 
		\left(\theta_9 -\beta _2 - \beta _3 + \gamma_2 - 1 \right),
		\label{eq:PDEsHexagon1} \\
\mathrm{PDE}_{13} &= 
	w_2 w_7 \brk[s]*{ w_4 \brk*{\theta _\Sigma + \alpha } - \theta _4 } \brk*{\theta _{2 6 7 8} +\beta _2 }
	+ w_3 w_7 \theta _2 \theta _4	
	-w_2 w_4 \theta _7 \left(\theta _{7 8 9}-\beta _3 + \gamma_2 -1\right), \\ 
\mathrm{PDE}_{14} &= 
	-w_4 w_8 \theta _5 \theta _7
	-w_5 w_7 \theta _4 \left(\theta _{4 5 6 7 9} +\gamma_2 -1\right) 
	+ w_4 w_5 w_7 \left(\theta _{3 4 5 6}+\beta _3\right) 
		\left(\theta _\Sigma + \alpha \right) , \\ 
\mathrm{PDE}_{15} &= 
	w_2 w_5 w_7 \theta _1 \theta _4 
	+w_1 w_6 w_7 \theta _2 \theta _4
	+w_2 w_4 w_8 \theta _1 \theta _7 
	+ w_1 w_2 w_4 w_7 \theta _{5 6 8} \left(\theta _\Sigma + \alpha \right)  , \\ 	
\mathrm{PDE}_{16} &= 
	\brk[s]*{ w_1 w_4 w_7 \brk*{ \theta _\Sigma + \alpha } 
		-w_5 w_7 \theta _4 -w_4 w_8 \theta _7} \left(\theta _{1 5 8 9}+\beta _1\right) 
	+ w_7 \brk[s]*{ w_2 w_4 \brk*{\theta _\Sigma + \alpha } - w_6 \theta _4}	
		\left(\theta _{2 6 7 8}+\beta _2\right) \nonumber \\
& \quad	-w_4 w_7 \brk*{ \theta _{1 2 3 5 6 8} + \gamma_1 -1 }	 \theta _{1 2 3 5 6 8} 	
  	+w_3 w_4 w_7 \left(\theta _{3 4 5 6}+\beta _3\right) 
		\left(\theta _\Sigma + \alpha \right) ,	\\
\mathrm{PDE}_{23} &= 
	 w_2 w_7 w_8 \brk[s]*{ w_1 \theta _5 + w_5 \theta _1 } \brk*{ \theta _{2 6 7 8} + \beta _2 }  
	+w_1 w_2 w_5 \brk[s]*{ w_8 \theta _7  \brk*{ \theta_{1 5 8 9} + \beta _1 } 
		+ w_7 \theta _8 \brk*{ \theta _8 -\alpha -\beta _3 + \gamma_1 + \gamma_2 -2} } \nonumber \\
& \quad	-w_1 w_3 w_7 w_8 \theta_2 \theta _5 
	-w_1 w_2 w_4 w_8 \theta_5 \theta _7
	-w_2 w_5 w_8 \theta _1 \theta _7 ,	\\	
\mathrm{PDE}_{24} &= 
	w_1 w_4 \brk*{ \theta _{5 6 8} -2 -\alpha +\gamma_1 +\gamma_2} \theta _5 
	+w_1 w_5 \brk*{\theta _{1 5 8 9} + \beta _1 } \theta _4
	+w_4 w_5 \brk*{\theta _{3 4 5 6} + \beta _3 } \theta _1
	-w_5 \theta _1 \theta _4  ,	\\	
\mathrm{PDE}_{25} &= 
	w_1 w_2 w_5 \brk*{ \theta _{1 5 8 9} + \beta _1 } \brk*{\theta _\Sigma + \alpha }
	-w_2 w_5 \brk{\theta _{1 2 3 5 8} -1 +\gamma_1 } \theta _1
	-w_1 w_6 \theta _2 \theta _5 ,	\\	
\mathrm{PDE}_{26} &= 
	w_1 w_5 \brk*{ \theta _{1 5 8 9}+\beta _1 } \theta _{2 3 6} 
	- \brk[s]*{w_2 w_5 \theta _1 + w_1 w_6 \theta _5} \left(\theta _{2 6 7 8}+\beta _2\right) 
	-w_3 w_5 \left(\theta _{3 4 5 6}+\beta _3\right) \theta _1 ,	\\	
\mathrm{PDE}_{34} &= 
	w_2 w_6 w_7	\brk[s]*{ w_5 \theta_4 + w_4  \theta _5 } \brk*{\theta _{2678}+\beta _2} 
	+w_4 w_5 w_6 w_7  \brk*{\theta _{3456}+\beta _3} \theta _2
	-w_5 w_6 w_7 \theta _2 \theta _4 \nonumber \\
& \quad +w_2 w_4 w_5 w_7 \brk*{\theta _6 -2 -\alpha -\beta _1 +\gamma_1 +\gamma_2} \theta _6 
	-w_1 w_4 w_6 w_7 \theta _2 \theta _5
	-w_2 w_4 w_6 w_9 \theta _5 \theta _7,	\\	
\mathrm{PDE}_{35} &= 
	w_2 w_7 \brk[s]*{ w_1 \brk*{\theta _\Sigma + \alpha } - \theta _1 } 
		\brk*{\theta _{2678}+\beta _2} 
	-w_1 w_7 \brk*{\theta _{236} -1 -\beta _1+ \gamma_1} \theta _2
	+w_2 w_9 \theta _1 \theta _7 ,	\\	
\mathrm{PDE}_{36} &=  w_2 w_9 \brk*{\theta _{1589}+\beta _1} \theta _7
	+w_2 w_7 \brk*{\theta _{2678}+\beta _2} \brk*{\theta _3-\theta _9}
	-w_3 w_7 \brk*{\theta _{3456}+\beta _3} \theta _2 ,	\\	
\mathrm{PDE}_{45} &= 
	w_3 w_4 w_5 \brk[s]*{w_1 w_2 \brk*{\theta _\Sigma + \alpha}
		-w_2 \theta _1 -w_1 \theta_2 } \brk*{\theta _{3456}+\beta _3} 
	-w_1 w_2 w_4 w_5 \brk*{ \theta _3 - 1 - \beta _1 -\beta _2+\gamma_1} \theta _3 \nonumber \\
& \quad +w_2 w_3 w_5 w_9 \theta _1 \theta _4 
	+w_1 w_3 w_5 w_7 \theta _2 \theta _4 
	+w_1 w_3 w_4 w_8 \theta _2 \theta _5 ,	\\	
\mathrm{PDE}_{46} &= 
	w_5 w_9 \brk*{\theta _{1589}+\beta _1} \theta _4
	+\brk[s]*{w_5 w_7 \theta _4 +w_4 w_8 \theta _5 } \brk*{\theta _{2678}+\beta _2}
	-w_4 w_5 \brk*{\theta _{3456}+\beta _3} \theta _{789} ,	\\	
\mathrm{PDE}_{56} &= 
	\brk[s]*{ w_1 w_2 w_4 \brk*{\theta _\Sigma + \alpha }
		-w_2 w_5 \theta _1 - w_1 w_6 \theta _2} \brk*{\theta _{3456}+\beta _3}
	+ w_2 \brk[s]*{ w_1 w_7 \brk*{\theta _\Sigma + \alpha} -\theta _1 w_2 w_8 }
		\brk*{\theta _{2678}+\beta _2} \nonumber \\
& \quad + w_1 w_2 w_9  \brk*{\theta _{1589}+\beta _1} \brk*{\theta _\Sigma + \alpha }
	-w_1 w_2 \theta _{456789} \brk*{\theta _{456789}-1+\gamma_2} . 
	\label{eq:PDEsHexagon15}			
\end{align}
\end{widetext}
\end{allowdisplaybreaks}

\section{Hexagon Recurrences}
\label{sec:RecEqsHex}
We introduce the shift operator $r_{m_1,\dots,m_9}$ with $m_k=\pm k$ or $m_k$ absent otherwise, which acts on the coefficient function $f_{n_1\dots n_9}$ and shifts the respective index $k$ by $\pm 1$ or $0$, respectively, e.g.\
\begin{align}
&r_{1,-3} \, f_{n_1n_2n_3n_4n_5n_6n_7n_8n_9}
\nonumber\\
&\qquad=f_{n_1+1,n_2,n_3-1,n_4n_5n_6n_7n_8n_9}.
\end{align}
With this notation and the shorthands $M_j$ defined in \eqref{eq:MHexa}, the recurrence equations for the hexagon function $f$ read 
\begin{equation}
\text{RE}_{jk}\, f_{n_1\dots n_9}=0,
\qquad
1\leq j<k\leq 6,
\end{equation}
 with the recurrence operators
 \begin{allowdisplaybreaks}
 \begin{widetext} 
\begin{align}
\text{RE}_{12}=
&-n_3 r_{1,-3,4}-n_2 r_{1,-2,7}-n_6 r_{5,-6,7}
-\left(2 \alpha +M_1\right) \left(2 \beta_1+M_2\right)+\left(r_4+r_7\right) \left(2 \beta _1+M_2\right)
   \nonumber\\
   &+r_9 \left(-\beta _2-\beta _3+\gamma_2+n_9\right),
\\
\text{RE}_{13}=
&n_3 r_{2,-3,4}+\left(2 \alpha +M_1\right) \left(2 \beta _2+M_3\right)
-r_4 \left(2 \beta
   _2+M_3\right)-r_7 \left(-\beta _3+\gamma _2+n_7+n_8+n_9\right),
   \\
\text{RE}_{14}=
&-n_8 r_{5,7,-8}+\left(2 \alpha +M_1\right) \left(2 \beta _3+M_4\right)
-r_4 \left(\gamma
   _2+n_4+n_5+n_6+n_7+n_9\right),
   \\
  \text{RE}_{15}=
  &n_5 r_{1,4,-5}+n_8 r_{1,7,-8}+n_6 r_{2,4,-6}
+\left(n_5+n_6+n_8\right) \left(2 \alpha
   +M_1\right),
   \\
  \text{RE}_{16}=
  &-n_5 r_{4,-5} \left(2 \beta _1+M_2-1\right)
-n_8 r_{7,-8} \left(2 \beta
   _1+M_2-1\right)-n_6 r_{4,-6} \left(2 \beta _2+M_3-1\right)
   \nonumber\\
&-\left(\gamma _1+M_5\right) \left(2 \gamma
   _1+M_5-1\right)+n_1 r_{-1} \left(2 \alpha +M_1-1\right) \left(2 \beta _1+M_2-1\right)
   \\
   &+n_2 r_{-2}
   \left(2 \alpha +M_1-1\right) \left(2 \beta _2+M_3-1\right)
+n_3 r_{-3} \left(2 \alpha +M_1-1\right)
   \left(2 \beta _3+M_4-1\right),
\nonumber   \\
   \text{RE}_{23}=&-n_4 r_{-4,5,7}-n_3 r_{2,-3,5}-r_{1,7}
+r_7 \left(2 \beta_1+M_2\right)+\left(r_1+r_5\right) \left(2 \beta _2+M_3\right)
\nonumber\\
&+r_8 \left(-\alpha -\beta _3+\gamma_1+\gamma _2+n_8-1\right),
\\
\text{RE}_{24}=&-r_{1,4}+r_4 \left(2 \beta _1+M_2\right)+r_1 \left(2 \beta _3+M_4\right)+r_5
   \left(-\alpha +\gamma _1+\gamma _2+n_5+n_6+n_8-1\right),
   \\
   \text{RE}_{25}=
   &-n_6 r_{2,5,-6}+\left(2 \alpha +M_1\right) \left(2 \beta _1+M_2\right)-r_1 \left(\gamma
   _1+n_1+n_2+n_3+n_5+n_8\right),
   \\
   \text{RE}_{26}=
   &-n_3 r_{1,-3} \left(2 \beta _3+M_4-1\right)-\left(2 \beta _2+M_3-1\right) \left(n_2
   r_{1,-2}+n_6 r_{5,-6}\right)+\left(n_2+n_3+n_6\right) \left(2 \beta _1+M_2\right),
   \\
   \text{RE}_{34}=
   &-n_1 r_{-1,2,5}-n_9 r_{5,7,-9}-r_{2,4}+\left(r_4+r_5\right) \left(2 \beta
   _2+M_3\right)+r_6 \left(-\alpha -\beta _1+\gamma _1+\gamma _2+n_6-1\right)
   \nonumber\\
   &+r_2 \left(\beta
   _3+n_6\right)+\left(n_3+n_4+n_5\right) r_2,
   \\
   \text{RE}_{35}=
   &n_9 r_{1,7,-9}+\left(2 \alpha +M_1\right) \left(2 \beta _2+M_3\right)-r_1 \left(2 \beta
   _2+M_3\right)-r_2 \left(-\beta _1+\gamma _1+n_2+n_3+n_6\right),
   \\
   \text{RE}_{36}=
   &-n_3 r_{2,-3} \left(2 \beta _3+M_4-1\right)+n_9 r_{7,-9} \left(2 \beta
   _1+M_2-1\right)+\left(n_3-n_9\right) \left(2 \beta _2+M_3\right),
   \\
   \text{RE}_{45}=
   &n_9 r_{1,4,-9}+n_7 r_{2,4,-7}+n_8 r_{2,5,-8}+\left(2 \alpha +M_1\right) \left(2 \beta
   _3+M_4\right)-\left(r_1+r_2\right) \left(2 \beta _3+M_4\right)
   \nonumber\\
   &+r_3 \left(\beta _1+\beta _2-\gamma
   _1-n_3\right),
   \\
   \text{RE}_{46}=
   &n_9 r_{4,-9} \left(2 \beta _1+M_2-1\right)+\left(2 \beta _2+M_3-1\right) \left(n_7
   r_{4,-7}+n_8 r_{5,-8}\right)-\left(n_7+n_8+n_9\right) \left(2 \beta _3+M_4\right),
   \\
   \text{RE}_{56}=
   &-n_8 r_{1,-8} \left(2 \beta _2+M_3-1\right)-n_5 r_{1,-5} \left(2 \beta
   _3+M_4-1\right)-n_6 r_{2,-6} \left(2 \beta _3+M_4-1\right)
\nonumber   \\
   &-\left(\gamma _2+M_6\right) \left(2 \gamma
   _2+M_6-1\right)+n_9 r_{-9} \left(2 \alpha +M_1-1\right) \left(2 \beta _1+M_2-1\right)
   \\
   &+n_7 r_{-7}
   \left(2 \alpha +M_1-1\right) \left(2 \beta _2+M_3-1\right)+n_4 r_{-4} \left(2 \alpha +M_1-1\right)
   \left(2 \beta _3+M_4-1\right).
   \nonumber
\end{align}
\end{widetext}
\end{allowdisplaybreaks}

\section{Useful Identities}
\label{sec:Identities}

We note Euler's reflection identity for the Gamma function
\begin{equation}
\Gamma_{1-z}\Gamma_z=\frac{\pi}{\sin \pi z},
\qquad
\text{for}\,\, z\notin \mathbb{Z},
\label{eq:ReflId1}
\end{equation}
and the resulting relation
\begin{equation}
\Gamma_{z-n}=(-1)^{n-1} \frac{\Gamma_{-z}\Gamma_{z+1}}{\Gamma_{n-z+1}},
\qquad
\text{for}\,\, n\in \mathbb{Z},\,\, z\notin\mathbb{Z}.
\label{eq:ReflId2}
\end{equation}

The Appell hypergeometric function $F_4$ obeys the useful identity
\begin{align}
&\FF{4}{\alpha,\beta}{\gamma,\gamma'}{u,v} = 
	\frac{\Gamma_{\gamma'} \Gamma_{\beta - \alpha} }
		{\Gamma_{\beta} \Gamma_{\gamma' - \alpha} } (-v)^{-\alpha}
	\FF{4}{\alpha,1+\alpha-\gamma'}{\gamma,1+\alpha-\beta}{u/v,1/v} \nonumber \\
& \quad + \frac{\Gamma_{\gamma'} \Gamma_{\alpha - \beta} }
		{\Gamma_{\alpha} \Gamma_{\gamma' - \beta} } (-v)^{-\beta}
	\FF{4}{1+\beta-\gamma',\beta}{\gamma,1+\beta-\alpha}{u/v,1/v} .
	\label{eq:FuncIdF4}
\end{align}
Gau{\ss}' hypergeometric function ${}_2F_1(a,b,c;z)$ evaluated at $z=1$ obeys
\begin{equation}
{}_2F_1(a,b,c;1)=\frac{\Gamma_{1+a-b}\Gamma_{1+a/2}}{\Gamma_{1+a}\Gamma_{1+a/2-b}}.
\label{eq:GaussAt1}
\end{equation}
\section{Evaluation Parameters}
\label{sec:EvalParam}

In this appendix we list the evaluation parameters entering the definition \eqref{eq:lev1}
of the Yangian level-one generators for the different integrals considered.

The evaluation parameters for the $D$-dimensional box integral \eqref{eq:BoxIntD} with $j=1,\dots,4$ read
\begin{align}
s_j = \half \big(b+c+D,b+2c+d,c+d,0\big)_j . 
\label{eq:BoxEvalParam}
\end{align}
The double box PDEs given in \eqref{eq:PDEsDoubleBox} are obtained by using the evaluation parameters (cf.\ \cite{Chicherin:2017cns})
\begin{equation}
s_j= -\brk!{0,1,2,2,3,4}_j.
\end{equation}

Finally, the hexagon PDEs given in \eqref{eq:PDEsHexagon1}--\eqref{eq:PDEsHexagon15} are obtained with the evaluation parameters 
\begin{align}
s_j=\half\big(
&a-D,2a+b-D,2a+2b+c-D,
\nonumber\\
&D-d-2e-2f,D-e-2f,D-f\big)_j,
\label{eq:HexEvalParam}
\end{align}
where $j=1,\dots,6$.


\bibliography{YangianDoubleBox}

\begin{thebibliography}{46}%
\makeatletter
\providecommand \@ifxundefined [1]{%
 \@ifx{#1\undefined}
}%
\providecommand \@ifnum [1]{%
 \ifnum #1\expandafter \@firstoftwo
 \else \expandafter \@secondoftwo
 \fi
}%
\providecommand \@ifx [1]{%
 \ifx #1\expandafter \@firstoftwo
 \else \expandafter \@secondoftwo
 \fi
}%
\providecommand \natexlab [1]{#1}%
\providecommand \enquote  [1]{``#1''}%
\providecommand \bibnamefont  [1]{#1}%
\providecommand \bibfnamefont [1]{#1}%
\providecommand \citenamefont [1]{#1}%
\providecommand \href@noop [0]{\@secondoftwo}%
\providecommand \href [0]{\begingroup \@sanitize@url \@href}%
\providecommand \@href[1]{\@@startlink{#1}\@@href}%
\providecommand \@@href[1]{\endgroup#1\@@endlink}%
\providecommand \@sanitize@url [0]{\catcode `\\12\catcode `\$12\catcode
  `\&12\catcode `\#12\catcode `\^12\catcode `\_12\catcode `\%12\relax}%
\providecommand \@@startlink[1]{}%
\providecommand \@@endlink[0]{}%
\providecommand \url  [0]{\begingroup\@sanitize@url \@url }%
\providecommand \@url [1]{\endgroup\@href {#1}{\urlprefix }}%
\providecommand \urlprefix  [0]{URL }%
\providecommand \Eprint [0]{\href }%
\providecommand \doibase [0]{http://dx.doi.org/}%
\providecommand \selectlanguage [0]{\@gobble}%
\providecommand \bibinfo  [0]{\@secondoftwo}%
\providecommand \bibfield  [0]{\@secondoftwo}%
\providecommand \translation [1]{[#1]}%
\providecommand \BibitemOpen [0]{}%
\providecommand \bibitemStop [0]{}%
\providecommand \bibitemNoStop [0]{.\EOS\space}%
\providecommand \EOS [0]{\spacefactor3000\relax}%
\providecommand \BibitemShut  [1]{\csname bibitem#1\endcsname}%
\let\auto@bib@innerbib\@empty
\bibitem [{\citenamefont {Chicherin}\ \emph
  {et~al.}(2018{\natexlab{a}})\citenamefont {Chicherin}, \citenamefont
  {Kazakov}, \citenamefont {Loebbert}, \citenamefont {M{\"u}ller},\ and\
  \citenamefont {Zhong}}]{Chicherin:2017cns}%
  \BibitemOpen
  \bibfield  {author} {\bibinfo {author} {\bibfnamefont {D.}~\bibnamefont
  {Chicherin}}, \bibinfo {author} {\bibfnamefont {V.}~\bibnamefont {Kazakov}},
  \bibinfo {author} {\bibfnamefont {F.}~\bibnamefont {Loebbert}}, \bibinfo
  {author} {\bibfnamefont {D.}~\bibnamefont {M{\"u}ller}}, \ and\ \bibinfo
  {author} {\bibfnamefont {D.-l.}\ \bibnamefont {Zhong}},\ }\href {\doibase
  10.1007/JHEP05(2018)003} {\bibfield  {journal} {\bibinfo  {journal} {JHEP}\
  }\textbf {\bibinfo {volume} {05}},\ \bibinfo {pages} {003} (\bibinfo {year}
  {2018}{\natexlab{a}})},\ \Eprint {http://arxiv.org/abs/1704.01967}
  {arXiv:1704.01967 [hep-th]} \BibitemShut {NoStop}%
\bibitem [{\citenamefont {Chicherin}\ \emph {et~al.}(2017)\citenamefont
  {Chicherin}, \citenamefont {Kazakov}, \citenamefont {Loebbert}, \citenamefont
  {M{\"u}ller},\ and\ \citenamefont {Zhong}}]{Chicherin:2017frs}%
  \BibitemOpen
  \bibfield  {author} {\bibinfo {author} {\bibfnamefont {D.}~\bibnamefont
  {Chicherin}}, \bibinfo {author} {\bibfnamefont {V.}~\bibnamefont {Kazakov}},
  \bibinfo {author} {\bibfnamefont {F.}~\bibnamefont {Loebbert}}, \bibinfo
  {author} {\bibfnamefont {D.}~\bibnamefont {M{\"u}ller}}, \ and\ \bibinfo
  {author} {\bibfnamefont {D.-l.}\ \bibnamefont {Zhong}},\ }\href {\doibase
  10.1103/PhysRevD.96.121901} {\bibfield  {journal} {\bibinfo  {journal} {Phys.
  Rev.}\ }\textbf {\bibinfo {volume} {D96}},\ \bibinfo {pages} {121901}
  (\bibinfo {year} {2017})},\ \Eprint {http://arxiv.org/abs/1708.00007}
  {arXiv:1708.00007 [hep-th]} \BibitemShut {NoStop}%
\bibitem [{\citenamefont {G{\"u}rdogan}\ and\ \citenamefont
  {Kazakov}(2016)}]{Gurdogan:2015csr}%
  \BibitemOpen
  \bibfield  {author} {\bibinfo {author} {\bibfnamefont {{\"O}.}~\bibnamefont
  {G{\"u}rdogan}}\ and\ \bibinfo {author} {\bibfnamefont {V.}~\bibnamefont
  {Kazakov}},\ }\href {\doibase 10.1103/PhysRevLett.117.201602,
  10.1103/PhysRevLett.117.259903} {\bibfield  {journal} {\bibinfo  {journal}
  {Phys. Rev. Lett.}\ }\textbf {\bibinfo {volume} {117}},\ \bibinfo {pages}
  {201602} (\bibinfo {year} {2016})},\ \bibinfo {note} {[Addendum: Phys. Rev.
  Lett.117,no.25,259903(2016)]},\ \Eprint {http://arxiv.org/abs/1512.06704}
  {arXiv:1512.06704 [hep-th]} \BibitemShut {NoStop}%
\bibitem [{\citenamefont {Zamolodchikov}(1980)}]{Zamolodchikov:1980mb}%
  \BibitemOpen
  \bibfield  {author} {\bibinfo {author} {\bibfnamefont {A.~B.}\ \bibnamefont
  {Zamolodchikov}},\ }\href {\doibase 10.1016/0370-2693(80)90547-X} {\bibfield
  {journal} {\bibinfo  {journal} {Phys. Lett.}\ }\textbf {\bibinfo {volume}
  {97B}},\ \bibinfo {pages} {63} (\bibinfo {year} {1980})}\BibitemShut
  {NoStop}%
\bibitem [{Note1()}]{Note1}%
  \BibitemOpen
  \bibinfo {note} {Though we will work in the dual momentum space throughout
  this paper, we refer to the respective Feynman diagrams in the original
  momentum space, i.e.\ we speak of the box instead of the cross integral. The
  hat over the integral symbol $\protect \mathaccentV {hat}05EI$ denotes the
  integral with undeformed propagator powers as opposed to the symbol $I$ used
  later for generic propagator powers.}\BibitemShut {Stop}%
\bibitem [{\citenamefont {Bourjaily}\ \emph {et~al.}(2018)\citenamefont
  {Bourjaily}, \citenamefont {McLeod}, \citenamefont {Spradlin}, \citenamefont
  {von Hippel},\ and\ \citenamefont {Wilhelm}}]{Bourjaily:2017bsb}%
  \BibitemOpen
  \bibfield  {author} {\bibinfo {author} {\bibfnamefont {J.~L.}\ \bibnamefont
  {Bourjaily}}, \bibinfo {author} {\bibfnamefont {A.~J.}\ \bibnamefont
  {McLeod}}, \bibinfo {author} {\bibfnamefont {M.}~\bibnamefont {Spradlin}},
  \bibinfo {author} {\bibfnamefont {M.}~\bibnamefont {von Hippel}}, \ and\
  \bibinfo {author} {\bibfnamefont {M.}~\bibnamefont {Wilhelm}},\ }\href
  {\doibase 10.1103/PhysRevLett.120.121603} {\bibfield  {journal} {\bibinfo
  {journal} {Phys. Rev. Lett.}\ }\textbf {\bibinfo {volume} {120}},\ \bibinfo
  {pages} {121603} (\bibinfo {year} {2018})},\ \Eprint
  {http://arxiv.org/abs/1712.02785} {arXiv:1712.02785 [hep-th]} \BibitemShut
  {NoStop}%
\bibitem [{\citenamefont {Adams}\ \emph {et~al.}(2018)\citenamefont {Adams},
  \citenamefont {Chaubey},\ and\ \citenamefont {Weinzierl}}]{Adams:2018bsn}%
  \BibitemOpen
  \bibfield  {author} {\bibinfo {author} {\bibfnamefont {L.}~\bibnamefont
  {Adams}}, \bibinfo {author} {\bibfnamefont {E.}~\bibnamefont {Chaubey}}, \
  and\ \bibinfo {author} {\bibfnamefont {S.}~\bibnamefont {Weinzierl}},\ }\href
  {\doibase 10.1103/PhysRevLett.121.142001} {\bibfield  {journal} {\bibinfo
  {journal} {Phys. Rev. Lett.}\ }\textbf {\bibinfo {volume} {121}},\ \bibinfo
  {pages} {142001} (\bibinfo {year} {2018})},\ \Eprint
  {http://arxiv.org/abs/1804.11144} {arXiv:1804.11144 [hep-ph]} \BibitemShut
  {NoStop}%
\bibitem [{\citenamefont {Caron-Huot}\ and\ \citenamefont
  {Larsen}(2012)}]{CaronHuot:2012ab}%
  \BibitemOpen
  \bibfield  {author} {\bibinfo {author} {\bibfnamefont {S.}~\bibnamefont
  {Caron-Huot}}\ and\ \bibinfo {author} {\bibfnamefont {K.~J.}\ \bibnamefont
  {Larsen}},\ }\href {\doibase 10.1007/JHEP10(2012)026} {\bibfield  {journal}
  {\bibinfo  {journal} {JHEP}\ }\textbf {\bibinfo {volume} {10}},\ \bibinfo
  {pages} {026} (\bibinfo {year} {2012})},\ \Eprint
  {http://arxiv.org/abs/1205.0801} {arXiv:1205.0801 [hep-ph]} \BibitemShut
  {NoStop}%
\bibitem [{\citenamefont {Paulos}\ \emph {et~al.}(2012)\citenamefont {Paulos},
  \citenamefont {Spradlin},\ and\ \citenamefont {Volovich}}]{Paulos:2012nu}%
  \BibitemOpen
  \bibfield  {author} {\bibinfo {author} {\bibfnamefont {M.~F.}\ \bibnamefont
  {Paulos}}, \bibinfo {author} {\bibfnamefont {M.}~\bibnamefont {Spradlin}}, \
  and\ \bibinfo {author} {\bibfnamefont {A.}~\bibnamefont {Volovich}},\ }\href
  {\doibase 10.1007/JHEP08(2012)072} {\bibfield  {journal} {\bibinfo  {journal}
  {JHEP}\ }\textbf {\bibinfo {volume} {08}},\ \bibinfo {pages} {072} (\bibinfo
  {year} {2012})},\ \Eprint {http://arxiv.org/abs/1203.6362} {arXiv:1203.6362
  [hep-th]} \BibitemShut {NoStop}%
\bibitem [{\citenamefont {Dixon}\ \emph {et~al.}(2011)\citenamefont {Dixon},
  \citenamefont {Drummond},\ and\ \citenamefont {Henn}}]{Dixon:2011ng}%
  \BibitemOpen
  \bibfield  {author} {\bibinfo {author} {\bibfnamefont {L.~J.}\ \bibnamefont
  {Dixon}}, \bibinfo {author} {\bibfnamefont {J.~M.}\ \bibnamefont {Drummond}},
  \ and\ \bibinfo {author} {\bibfnamefont {J.~M.}\ \bibnamefont {Henn}},\
  }\href {\doibase 10.1007/JHEP06(2011)100} {\bibfield  {journal} {\bibinfo
  {journal} {JHEP}\ }\textbf {\bibinfo {volume} {06}},\ \bibinfo {pages} {100}
  (\bibinfo {year} {2011})},\ \Eprint {http://arxiv.org/abs/1104.2787}
  {arXiv:1104.2787 [hep-th]} \BibitemShut {NoStop}%
\bibitem [{\citenamefont {Del~Duca}\ \emph {et~al.}(2011)\citenamefont
  {Del~Duca}, \citenamefont {Dixon}, \citenamefont {Drummond}, \citenamefont
  {Duhr}, \citenamefont {Henn},\ and\ \citenamefont
  {Smirnov}}]{DelDuca:2011wh}%
  \BibitemOpen
  \bibfield  {author} {\bibinfo {author} {\bibfnamefont {V.}~\bibnamefont
  {Del~Duca}}, \bibinfo {author} {\bibfnamefont {L.~J.}\ \bibnamefont {Dixon}},
  \bibinfo {author} {\bibfnamefont {J.~M.}\ \bibnamefont {Drummond}}, \bibinfo
  {author} {\bibfnamefont {C.}~\bibnamefont {Duhr}}, \bibinfo {author}
  {\bibfnamefont {J.~M.}\ \bibnamefont {Henn}}, \ and\ \bibinfo {author}
  {\bibfnamefont {V.~A.}\ \bibnamefont {Smirnov}},\ }\href {\doibase
  10.1103/PhysRevD.84.045017} {\bibfield  {journal} {\bibinfo  {journal} {Phys.
  Rev.}\ }\textbf {\bibinfo {volume} {D84}},\ \bibinfo {pages} {045017}
  (\bibinfo {year} {2011})},\ \Eprint {http://arxiv.org/abs/1105.2011}
  {arXiv:1105.2011 [hep-th]} \BibitemShut {NoStop}%
\bibitem [{\citenamefont {Loebbert}\ and\ \citenamefont
  {Spiering}(2018)}]{Loebbert:2018lxk}%
  \BibitemOpen
  \bibfield  {author} {\bibinfo {author} {\bibfnamefont {F.}~\bibnamefont
  {Loebbert}}\ and\ \bibinfo {author} {\bibfnamefont {A.}~\bibnamefont
  {Spiering}},\ }\href {\doibase 10.1088/1751-8121/aae7ff} {\bibfield
  {journal} {\bibinfo  {journal} {J. Phys.}\ }\textbf {\bibinfo {volume}
  {A51}},\ \bibinfo {pages} {485202} (\bibinfo {year} {2018})},\ \Eprint
  {http://arxiv.org/abs/1805.11993} {arXiv:1805.11993 [hep-th]} \BibitemShut
  {NoStop}%
\bibitem [{Note2()}]{Note2}%
  \BibitemOpen
  \bibinfo {note} {Note that by inclusion of a non-trivial numerator factor we
  could make the integrals conformally invariant.}\BibitemShut {Stop}%
\bibitem [{\citenamefont {Chicherin}\ \emph {et~al.}(2013)\citenamefont
  {Chicherin}, \citenamefont {Derkachov},\ and\ \citenamefont
  {Isaev}}]{Chicherin:2012yn}%
  \BibitemOpen
  \bibfield  {author} {\bibinfo {author} {\bibfnamefont {D.}~\bibnamefont
  {Chicherin}}, \bibinfo {author} {\bibfnamefont {S.}~\bibnamefont
  {Derkachov}}, \ and\ \bibinfo {author} {\bibfnamefont {A.~P.}\ \bibnamefont
  {Isaev}},\ }\href {\doibase 10.1007/JHEP04(2013)020} {\bibfield  {journal}
  {\bibinfo  {journal} {JHEP}\ }\textbf {\bibinfo {volume} {04}},\ \bibinfo
  {pages} {020} (\bibinfo {year} {2013})},\ \Eprint
  {http://arxiv.org/abs/1206.4150} {arXiv:1206.4150 [math-ph]} \BibitemShut
  {NoStop}%
\bibitem [{Note3()}]{Note3}%
  \BibitemOpen
  \bibinfo {note} {In addition the definition requires the so-called Serre
  relations, see e.g.\ \cite {Loebbert:2016cdm} for more details. Invariance
  under the full level-zero (conformal) algebra and a single level-one
  generator guarantees invariance under the full Yangian algebra.}\BibitemShut
  {Stop}%
\bibitem [{Note4()}]{Note4}%
  \BibitemOpen
  \bibinfo {note} {Here one may wonder how the 15 degrees of freedom of the
  conformal group can lead to 15 invariance conditions, when it is clear that
  translations and dilatations do not contribute. The point is that the
  conformal group acts non-linearly on the $x_{ij} ^\mu / x_{ij}^2$ such that
  the linear counting of degrees of freedom does not work out.}\BibitemShut
  {Stop}%
\bibitem [{\citenamefont {Usyukina}\ and\ \citenamefont
  {Davydychev}(1993)}]{Usyukina:1992jd}%
  \BibitemOpen
  \bibfield  {author} {\bibinfo {author} {\bibfnamefont {N.~I.}\ \bibnamefont
  {Usyukina}}\ and\ \bibinfo {author} {\bibfnamefont {A.~I.}\ \bibnamefont
  {Davydychev}},\ }\href {\doibase 10.1016/0370-2693(93)91834-A} {\bibfield
  {journal} {\bibinfo  {journal} {Phys. Lett.}\ }\textbf {\bibinfo {volume}
  {B298}},\ \bibinfo {pages} {363} (\bibinfo {year} {1993})}\BibitemShut
  {NoStop}%
\bibitem [{Note5()}]{Note5}%
  \BibitemOpen
  \bibinfo {note} {Note that the Appell function $F_4$ is symmetric in $\alpha
  $ and $\beta $.}\BibitemShut {Stop}%
\bibitem [{\citenamefont {Appell}\ and\ \citenamefont
  {De~F{\'e}riet}(1926)}]{appell1926fonctions}%
  \BibitemOpen
  \bibfield  {author} {\bibinfo {author} {\bibfnamefont {P.}~\bibnamefont
  {Appell}}\ and\ \bibinfo {author} {\bibfnamefont {J.~K.}\ \bibnamefont
  {De~F{\'e}riet}},\ }\href@noop {} {\emph {\bibinfo {title} {Fonctions
  hyperg{\'e}om{\'e}triques et hypersph{\'e}riques: polynomes d'Hermite}}}\
  (\bibinfo  {publisher} {Gauthier-Villars},\ \bibinfo {year}
  {1926})\BibitemShut {NoStop}%
\bibitem [{\citenamefont {Boos}\ and\ \citenamefont
  {Davydychev}(1991)}]{Boos:1990rg}%
  \BibitemOpen
  \bibfield  {author} {\bibinfo {author} {\bibfnamefont {E.~E.}\ \bibnamefont
  {Boos}}\ and\ \bibinfo {author} {\bibfnamefont {A.~I.}\ \bibnamefont
  {Davydychev}},\ }\href {\doibase 10.1007/BF01016805} {\bibfield  {journal}
  {\bibinfo  {journal} {Theor. Math. Phys.}\ }\textbf {\bibinfo {volume}
  {89}},\ \bibinfo {pages} {1052} (\bibinfo {year} {1991})},\ \bibinfo {note}
  {[Teor. Mat. Fiz.89,56(1991)]}\BibitemShut {NoStop}%
\bibitem [{\citenamefont {Davydychev}(1992)}]{Davydychev:1992xr}%
  \BibitemOpen
  \bibfield  {author} {\bibinfo {author} {\bibfnamefont {A.~I.}\ \bibnamefont
  {Davydychev}},\ }\href@noop {} {\bibfield  {journal} {\bibinfo  {journal} {J.
  Phys.}\ }\textbf {\bibinfo {volume} {A25}},\ \bibinfo {pages} {5587}
  (\bibinfo {year} {1992})}\BibitemShut {NoStop}%
\bibitem [{Note6()}]{Note6}%
  \BibitemOpen
  \bibinfo {note} {Note that the identity \protect \eqref {eq:ReflId2} given in
  the appendix can be useful to bring the solution into this form.}\BibitemShut
  {Stop}%
\bibitem [{\citenamefont {Nandan}\ \emph {et~al.}(2013)\citenamefont {Nandan},
  \citenamefont {Paulos}, \citenamefont {Spradlin},\ and\ \citenamefont
  {Volovich}}]{Nandan:2013ip}%
  \BibitemOpen
  \bibfield  {author} {\bibinfo {author} {\bibfnamefont {D.}~\bibnamefont
  {Nandan}}, \bibinfo {author} {\bibfnamefont {M.~F.}\ \bibnamefont {Paulos}},
  \bibinfo {author} {\bibfnamefont {M.}~\bibnamefont {Spradlin}}, \ and\
  \bibinfo {author} {\bibfnamefont {A.}~\bibnamefont {Volovich}},\ }\href
  {\doibase 10.1007/JHEP05(2013)105} {\bibfield  {journal} {\bibinfo  {journal}
  {JHEP}\ }\textbf {\bibinfo {volume} {05}},\ \bibinfo {pages} {105} (\bibinfo
  {year} {2013})},\ \Eprint {http://arxiv.org/abs/1301.2500} {arXiv:1301.2500
  [hep-th]} \BibitemShut {NoStop}%
\bibitem [{\citenamefont {Srivastava}\ and\ \citenamefont
  {Daoust}(1969)}]{srivastava1969certain}%
  \BibitemOpen
  \bibfield  {author} {\bibinfo {author} {\bibfnamefont {H.}~\bibnamefont
  {Srivastava}}\ and\ \bibinfo {author} {\bibfnamefont {M.~C.}\ \bibnamefont
  {Daoust}},\ }\href@noop {} {\bibfield  {journal} {\bibinfo  {journal}
  {Proceedings of the Koninklijke Nederlandse Akademie van Wetenschappen Series
  A-Mathematical Sciences}\ }\textbf {\bibinfo {volume} {72}},\ \bibinfo
  {pages} {449} (\bibinfo {year} {1969})}\BibitemShut {NoStop}%
\bibitem [{\citenamefont {Srivastava}\ and\ \citenamefont
  {Karlsson}(1985)}]{srivastava1985multiple}%
  \BibitemOpen
  \bibfield  {author} {\bibinfo {author} {\bibfnamefont {H.~M.}\ \bibnamefont
  {Srivastava}}\ and\ \bibinfo {author} {\bibfnamefont {P.~W.}\ \bibnamefont
  {Karlsson}},\ }\href@noop {} {\emph {\bibinfo {title} {Multiple Gaussian
  hypergeometric series}}}\ (\bibinfo  {publisher} {Ellis Horwood},\ \bibinfo
  {year} {1985})\BibitemShut {NoStop}%
\bibitem [{\citenamefont {Passare}\ \emph {et~al.}(1994)\citenamefont
  {Passare}, \citenamefont {Tsikh},\ and\ \citenamefont
  {Zhdanov}}]{Passare1994}%
  \BibitemOpen
  \bibfield  {author} {\bibinfo {author} {\bibfnamefont {M.}~\bibnamefont
  {Passare}}, \bibinfo {author} {\bibfnamefont {A.}~\bibnamefont {Tsikh}}, \
  and\ \bibinfo {author} {\bibfnamefont {O.}~\bibnamefont {Zhdanov}},\
  }\enquote {\bibinfo {title} {A multidimensional jordan residue lemma with an
  application to mellin-barnes integrals},}\ in\ \href {\doibase
  10.1007/978-3-663-14196-9_8} {\emph {\bibinfo {booktitle} {Contributions to
  Complex Analysis and Analytic Geometry: Dedicated to Pierre Dolbeault}}},\
  \bibinfo {editor} {edited by\ \bibinfo {editor} {\bibfnamefont
  {H.}~\bibnamefont {Skoda}}\ and\ \bibinfo {editor} {\bibfnamefont {J.-M.}\
  \bibnamefont {Tr{\'e}preau}}}\ (\bibinfo  {publisher} {Vieweg+Teubner
  Verlag},\ \bibinfo {address} {Wiesbaden},\ \bibinfo {year} {1994})\ pp.\
  \bibinfo {pages} {233--241}\BibitemShut {NoStop}%
\bibitem [{\citenamefont {Zhdanov}\ and\ \citenamefont
  {Tsikh}(1998)}]{zhdanov1998studying}%
  \BibitemOpen
  \bibfield  {author} {\bibinfo {author} {\bibfnamefont {O.}~\bibnamefont
  {Zhdanov}}\ and\ \bibinfo {author} {\bibfnamefont {A.}~\bibnamefont
  {Tsikh}},\ }\href@noop {} {\bibfield  {journal} {\bibinfo  {journal}
  {Siberian Mathematical Journal}\ }\textbf {\bibinfo {volume} {39}},\ \bibinfo
  {pages} {245} (\bibinfo {year} {1998})}\BibitemShut {NoStop}%
\bibitem [{\citenamefont {Friot}\ and\ \citenamefont
  {Greynat}(2012)}]{Friot:2011ic}%
  \BibitemOpen
  \bibfield  {author} {\bibinfo {author} {\bibfnamefont {S.}~\bibnamefont
  {Friot}}\ and\ \bibinfo {author} {\bibfnamefont {D.}~\bibnamefont
  {Greynat}},\ }\href {\doibase 10.1063/1.3679686} {\bibfield  {journal}
  {\bibinfo  {journal} {J. Math. Phys.}\ }\textbf {\bibinfo {volume} {53}},\
  \bibinfo {pages} {023508} (\bibinfo {year} {2012})},\ \Eprint
  {http://arxiv.org/abs/1107.0328} {arXiv:1107.0328 [math-ph]} \BibitemShut
  {NoStop}%
\bibitem [{\citenamefont {Bargheer}\ \emph {et~al.}(2009)\citenamefont
  {Bargheer}, \citenamefont {Beisert}, \citenamefont {Galleas}, \citenamefont
  {Loebbert},\ and\ \citenamefont {McLoughlin}}]{Bargheer:2009qu}%
  \BibitemOpen
  \bibfield  {author} {\bibinfo {author} {\bibfnamefont {T.}~\bibnamefont
  {Bargheer}}, \bibinfo {author} {\bibfnamefont {N.}~\bibnamefont {Beisert}},
  \bibinfo {author} {\bibfnamefont {W.}~\bibnamefont {Galleas}}, \bibinfo
  {author} {\bibfnamefont {F.}~\bibnamefont {Loebbert}}, \ and\ \bibinfo
  {author} {\bibfnamefont {T.}~\bibnamefont {McLoughlin}},\ }\href {\doibase
  10.1088/1126-6708/2009/11/056} {\bibfield  {journal} {\bibinfo  {journal}
  {JHEP}\ }\textbf {\bibinfo {volume} {11}},\ \bibinfo {pages} {056} (\bibinfo
  {year} {2009})},\ \Eprint {http://arxiv.org/abs/0905.3738} {arXiv:0905.3738
  [hep-th]} \BibitemShut {NoStop}%
\bibitem [{\citenamefont {Chicherin}\ and\ \citenamefont
  {Sokatchev}(2018)}]{Chicherin:2017bxc}%
  \BibitemOpen
  \bibfield  {author} {\bibinfo {author} {\bibfnamefont {D.}~\bibnamefont
  {Chicherin}}\ and\ \bibinfo {author} {\bibfnamefont {E.}~\bibnamefont
  {Sokatchev}},\ }\href {\doibase 10.1007/JHEP04(2018)082} {\bibfield
  {journal} {\bibinfo  {journal} {JHEP}\ }\textbf {\bibinfo {volume} {04}},\
  \bibinfo {pages} {082} (\bibinfo {year} {2018})},\ \Eprint
  {http://arxiv.org/abs/1709.03511} {arXiv:1709.03511 [hep-th]} \BibitemShut
  {NoStop}%
\bibitem [{\citenamefont {Chicherin}\ \emph
  {et~al.}(2018{\natexlab{b}})\citenamefont {Chicherin}, \citenamefont {Henn},\
  and\ \citenamefont {Sokatchev}}]{Chicherin:2018ubl}%
  \BibitemOpen
  \bibfield  {author} {\bibinfo {author} {\bibfnamefont {D.}~\bibnamefont
  {Chicherin}}, \bibinfo {author} {\bibfnamefont {J.~M.}\ \bibnamefont {Henn}},
  \ and\ \bibinfo {author} {\bibfnamefont {E.}~\bibnamefont {Sokatchev}},\
  }\href {\doibase 10.1103/PhysRevLett.121.021602} {\bibfield  {journal}
  {\bibinfo  {journal} {Phys. Rev. Lett.}\ }\textbf {\bibinfo {volume} {121}},\
  \bibinfo {pages} {021602} (\bibinfo {year} {2018}{\natexlab{b}})},\ \Eprint
  {http://arxiv.org/abs/1804.03571} {arXiv:1804.03571 [hep-th]} \BibitemShut
  {NoStop}%
\bibitem [{\citenamefont {Broedel}\ \emph {et~al.}(2018)\citenamefont
  {Broedel}, \citenamefont {Duhr}, \citenamefont {Dulat},\ and\ \citenamefont
  {Tancredi}}]{Broedel:2017kkb}%
  \BibitemOpen
  \bibfield  {author} {\bibinfo {author} {\bibfnamefont {J.}~\bibnamefont
  {Broedel}}, \bibinfo {author} {\bibfnamefont {C.}~\bibnamefont {Duhr}},
  \bibinfo {author} {\bibfnamefont {F.}~\bibnamefont {Dulat}}, \ and\ \bibinfo
  {author} {\bibfnamefont {L.}~\bibnamefont {Tancredi}},\ }\href {\doibase
  10.1007/JHEP05(2018)093} {\bibfield  {journal} {\bibinfo  {journal} {JHEP}\
  }\textbf {\bibinfo {volume} {05}},\ \bibinfo {pages} {093} (\bibinfo {year}
  {2018})},\ \Eprint {http://arxiv.org/abs/1712.07089} {arXiv:1712.07089
  [hep-th]} \BibitemShut {NoStop}%
\bibitem [{\citenamefont {de~la Cruz}(2019)}]{delaCruz:2019skx}%
  \BibitemOpen
  \bibfield  {author} {\bibinfo {author} {\bibfnamefont {L.}~\bibnamefont
  {de~la Cruz}},\ }\href@noop {} {\  (\bibinfo {year} {2019})},\ \Eprint
  {http://arxiv.org/abs/1907.00507} {arXiv:1907.00507 [math-ph]} \BibitemShut
  {NoStop}%
\bibitem [{\citenamefont {Klausen}(2019)}]{Klausen:2019hrg}%
  \BibitemOpen
  \bibfield  {author} {\bibinfo {author} {\bibfnamefont {R.~P.}\ \bibnamefont
  {Klausen}},\ }\href@noop {} {\  (\bibinfo {year} {2019})},\ \Eprint
  {http://arxiv.org/abs/1910.08651} {arXiv:1910.08651 [hep-th]} \BibitemShut
  {NoStop}%
\bibitem [{\citenamefont {Feng}\ \emph {et~al.}(2019)\citenamefont {Feng},
  \citenamefont {Chang}, \citenamefont {Chen},\ and\ \citenamefont
  {Zhang}}]{Feng:2019bdx}%
  \BibitemOpen
  \bibfield  {author} {\bibinfo {author} {\bibfnamefont {T.-F.}\ \bibnamefont
  {Feng}}, \bibinfo {author} {\bibfnamefont {C.-H.}\ \bibnamefont {Chang}},
  \bibinfo {author} {\bibfnamefont {J.-B.}\ \bibnamefont {Chen}}, \ and\
  \bibinfo {author} {\bibfnamefont {H.-B.}\ \bibnamefont {Zhang}},\ }\href@noop
  {} {\  (\bibinfo {year} {2019})},\ \Eprint {http://arxiv.org/abs/1912.01726}
  {arXiv:1912.01726 [hep-th]} \BibitemShut {NoStop}%
\bibitem [{\citenamefont {Brown}\ and\ \citenamefont
  {Dupont}(2019)}]{Brown:2019jng}%
  \BibitemOpen
  \bibfield  {author} {\bibinfo {author} {\bibfnamefont {F.}~\bibnamefont
  {Brown}}\ and\ \bibinfo {author} {\bibfnamefont {C.}~\bibnamefont {Dupont}},\
  }\href@noop {} {\  (\bibinfo {year} {2019})},\ \Eprint
  {http://arxiv.org/abs/1907.06603} {arXiv:1907.06603 [math.AG]} \BibitemShut
  {NoStop}%
\bibitem [{\citenamefont {Abreu}\ \emph {et~al.}(2019)\citenamefont {Abreu},
  \citenamefont {Britto}, \citenamefont {Duhr}, \citenamefont {Gardi},\ and\
  \citenamefont {Matthew}}]{Abreu:2019wzk}%
  \BibitemOpen
  \bibfield  {author} {\bibinfo {author} {\bibfnamefont {S.}~\bibnamefont
  {Abreu}}, \bibinfo {author} {\bibfnamefont {R.}~\bibnamefont {Britto}},
  \bibinfo {author} {\bibfnamefont {C.}~\bibnamefont {Duhr}}, \bibinfo {author}
  {\bibfnamefont {E.}~\bibnamefont {Gardi}}, \ and\ \bibinfo {author}
  {\bibfnamefont {J.}~\bibnamefont {Matthew}},\ }\href@noop {} {\  (\bibinfo
  {year} {2019})},\ \Eprint {http://arxiv.org/abs/1910.08358} {arXiv:1910.08358
  [hep-th]} \BibitemShut {NoStop}%
\bibitem [{\citenamefont {Isachenkov}\ and\ \citenamefont
  {Schomerus}(2016)}]{Isachenkov:2016gim}%
  \BibitemOpen
  \bibfield  {author} {\bibinfo {author} {\bibfnamefont {M.}~\bibnamefont
  {Isachenkov}}\ and\ \bibinfo {author} {\bibfnamefont {V.}~\bibnamefont
  {Schomerus}},\ }\href {\doibase 10.1103/PhysRevLett.117.071602} {\bibfield
  {journal} {\bibinfo  {journal} {Phys. Rev. Lett.}\ }\textbf {\bibinfo
  {volume} {117}},\ \bibinfo {pages} {071602} (\bibinfo {year} {2016})},\
  \Eprint {http://arxiv.org/abs/1602.01858} {arXiv:1602.01858 [hep-th]}
  \BibitemShut {NoStop}%
\bibitem [{\citenamefont {Caetano}\ \emph {et~al.}(2018)\citenamefont
  {Caetano}, \citenamefont {G{\"u}rdogan},\ and\ \citenamefont
  {Kazakov}}]{Caetano:2016ydc}%
  \BibitemOpen
  \bibfield  {author} {\bibinfo {author} {\bibfnamefont {J.}~\bibnamefont
  {Caetano}}, \bibinfo {author} {\bibfnamefont {{\"O}.}~\bibnamefont
  {G{\"u}rdogan}}, \ and\ \bibinfo {author} {\bibfnamefont {V.}~\bibnamefont
  {Kazakov}},\ }\href {\doibase 10.1007/JHEP03(2018)077} {\bibfield  {journal}
  {\bibinfo  {journal} {JHEP}\ }\textbf {\bibinfo {volume} {03}},\ \bibinfo
  {pages} {077} (\bibinfo {year} {2018})},\ \Eprint
  {http://arxiv.org/abs/1612.05895} {arXiv:1612.05895 [hep-th]} \BibitemShut
  {NoStop}%
\bibitem [{\citenamefont {Kazakov}\ \emph {et~al.}(2019)\citenamefont
  {Kazakov}, \citenamefont {Olivucci},\ and\ \citenamefont
  {Preti}}]{Kazakov:2018gcy}%
  \BibitemOpen
  \bibfield  {author} {\bibinfo {author} {\bibfnamefont {V.}~\bibnamefont
  {Kazakov}}, \bibinfo {author} {\bibfnamefont {E.}~\bibnamefont {Olivucci}}, \
  and\ \bibinfo {author} {\bibfnamefont {M.}~\bibnamefont {Preti}},\ }\href
  {\doibase 10.1007/JHEP06(2019)078} {\bibfield  {journal} {\bibinfo  {journal}
  {JHEP}\ }\textbf {\bibinfo {volume} {06}},\ \bibinfo {pages} {078} (\bibinfo
  {year} {2019})},\ \Eprint {http://arxiv.org/abs/1901.00011} {arXiv:1901.00011
  [hep-th]} \BibitemShut {NoStop}%
\bibitem [{\citenamefont {Dirac}(1936)}]{Dirac:1936fq}%
  \BibitemOpen
  \bibfield  {author} {\bibinfo {author} {\bibfnamefont {P.~A.~M.}\
  \bibnamefont {Dirac}},\ }\href {\doibase 10.2307/1968455} {\bibfield
  {journal} {\bibinfo  {journal} {Annals Math.}\ }\textbf {\bibinfo {volume}
  {37}},\ \bibinfo {pages} {429} (\bibinfo {year} {1936})}\BibitemShut
  {NoStop}%
\bibitem [{\citenamefont {Dorn}\ \emph {et~al.}(2014)\citenamefont {Dorn},
  \citenamefont {M{\"u}nkler},\ and\ \citenamefont {Spielvogel}}]{Dorn:2012cn}%
  \BibitemOpen
  \bibfield  {author} {\bibinfo {author} {\bibfnamefont {H.}~\bibnamefont
  {Dorn}}, \bibinfo {author} {\bibfnamefont {H.}~\bibnamefont {M{\"u}nkler}}, \
  and\ \bibinfo {author} {\bibfnamefont {C.}~\bibnamefont {Spielvogel}},\
  }\href {\doibase 10.1134/S1063779614040066} {\bibfield  {journal} {\bibinfo
  {journal} {Phys. Part. Nucl.}\ }\textbf {\bibinfo {volume} {45}},\ \bibinfo
  {pages} {692} (\bibinfo {year} {2014})},\ \Eprint
  {http://arxiv.org/abs/1211.5537} {arXiv:1211.5537 [hep-th]} \BibitemShut
  {NoStop}%
\bibitem [{\citenamefont {Hodges}(2013)}]{Hodges:2010kq}%
  \BibitemOpen
  \bibfield  {author} {\bibinfo {author} {\bibfnamefont {A.}~\bibnamefont
  {Hodges}},\ }\href {\doibase 10.1007/JHEP08(2013)051} {\bibfield  {journal}
  {\bibinfo  {journal} {JHEP}\ }\textbf {\bibinfo {volume} {08}},\ \bibinfo
  {pages} {051} (\bibinfo {year} {2013})},\ \Eprint
  {http://arxiv.org/abs/1004.3323} {arXiv:1004.3323 [hep-th]} \BibitemShut
  {NoStop}%
\bibitem [{\citenamefont {Simmons-Duffin}(2014)}]{SimmonsDuffin:2012uy}%
  \BibitemOpen
  \bibfield  {author} {\bibinfo {author} {\bibfnamefont {D.}~\bibnamefont
  {Simmons-Duffin}},\ }\href {\doibase 10.1007/JHEP04(2014)146} {\bibfield
  {journal} {\bibinfo  {journal} {JHEP}\ }\textbf {\bibinfo {volume} {04}},\
  \bibinfo {pages} {146} (\bibinfo {year} {2014})},\ \Eprint
  {http://arxiv.org/abs/1204.3894} {arXiv:1204.3894 [hep-th]} \BibitemShut
  {NoStop}%
\bibitem [{\citenamefont {Bourjaily}\ \emph {et~al.}(2019)\citenamefont
  {Bourjaily}, \citenamefont {Dulat},\ and\ \citenamefont
  {Panzer}}]{Bourjaily:2019jrk}%
  \BibitemOpen
  \bibfield  {author} {\bibinfo {author} {\bibfnamefont {J.~L.}\ \bibnamefont
  {Bourjaily}}, \bibinfo {author} {\bibfnamefont {F.}~\bibnamefont {Dulat}}, \
  and\ \bibinfo {author} {\bibfnamefont {E.}~\bibnamefont {Panzer}},\ }\href
  {\doibase 10.1016/j.nuclphysb.2019.03.022} {\bibfield  {journal} {\bibinfo
  {journal} {Nucl. Phys.}\ }\textbf {\bibinfo {volume} {B942}},\ \bibinfo
  {pages} {251} (\bibinfo {year} {2019})},\ \Eprint
  {http://arxiv.org/abs/1901.02887} {arXiv:1901.02887 [hep-th]} \BibitemShut
  {NoStop}%
\bibitem [{\citenamefont {Loebbert}(2016)}]{Loebbert:2016cdm}%
  \BibitemOpen
  \bibfield  {author} {\bibinfo {author} {\bibfnamefont {F.}~\bibnamefont
  {Loebbert}},\ }\href {\doibase 10.1088/1751-8113/49/32/323002} {\bibfield
  {journal} {\bibinfo  {journal} {J. Phys.}\ }\textbf {\bibinfo {volume}
  {A49}},\ \bibinfo {pages} {323002} (\bibinfo {year} {2016})},\ \Eprint
  {http://arxiv.org/abs/1606.02947} {arXiv:1606.02947 [hep-th]} \BibitemShut
  {NoStop}%
\end{thebibliography}%

\end{document}